\documentclass[preprint,3p,12pt,authoryear]{elsarticle}

\usepackage{url}
\usepackage{multirow}
\usepackage{graphicx}
\usepackage{dcolumn}
\usepackage{bm}
\usepackage{wrapfig}
\usepackage{float}
\usepackage{threeparttable}
\usepackage{footnote}
\usepackage[fleqn]{amsmath}
\usepackage{amssymb}
\usepackage{latexsym}
\usepackage{epsfig}
\usepackage{amsbsy}
\usepackage{array}
\usepackage{amssymb}
\usepackage{bm}
\usepackage{diagbox}
\usepackage{booktabs}
\usepackage{tikz}
\usepackage{subfigure}
\usepackage{epstopdf}
\usepackage[colorlinks=true,linkcolor=blue,anchorcolor=blue,citecolor=blue,urlcolor=blue]{hyperref}
\usepackage{natbib}
\setcitestyle{authortitle,open={(},close={)}}
\usetikzlibrary{shapes.geometric, arrows}

\begin{document}

\title{Calibration of the Instrumental Response of \emph{Insight-HXMT}/HE CsI Detectors for Gamma-Ray Monitoring}

\author[1,2]{Qi Luo\corref{cor1}}
\ead{luoqi@ihep.ac.cn}
\author[1]{Jin-Yuan Liao\corref{cor1}}
\ead{liaojinyuan@ihep.ac.cn}
\cortext[cor1]{Corresponding author}
\author[1]{Xu-Fang Li}
\author[1]{Gang Li}
\author[1]{Juan Zhang}
\author[1]{Cong-Zhan Liu}
\author[1]{Xiao-Bo Li}
\author[1]{Yue Zhu}
\author[1]{Cheng-Kui Li}
\author[1]{Yue Huang}
\author[1]{Ming-Yu Ge}
\author[1]{Yu-Peng Xu}
\author[1]{Zheng-Wei Li}
\author[1,2]{Ce Cai}
\author[1,2]{Shuo Xiao}
\author[1,4]{Qi-Bin Yi}
\author[1]{Yi-Fei Zhang}
\author[1]{Shao-Lin Xiong}
\author[1]{Shu Zhang}
\author[1,2,3]{Shuang-Nan Zhang}

\address[1]{Key Laboratory of Particle Astrophysics, Institute of High Energy Physics,
Chinese Academy of Sciences, Beijing 100049, China}
\address[2]{University of Chinese Academy of Sciences, Chinese Academy of Sciences, Beijing 100049, China}
\address[3]{National Astronomical Observatories, Chinese Academy of Sciences, Beijing, 100012, China}
\address[4]{School of Physics and Optoelectronics, Xiangtan University, Xiangtan, Hunan, 411105, China}

\begin{abstract}
The CsI detectors of the High Energy X-ray Telescope of the \emph{Hard X-ray Modulation Telescope} (\emph{HXMT}/CsI) can be used for gamma-ray
all sky monitoring and searching for the electromagnetic counterpart of gravitational wave source. The instrumental responses are mainly
obtained by Monte Carlo simulation with the Geant4 tool and the mass model of both the satellite and all the payloads, which is
updated and tested with the Crab pulse emission in various incident directions. Both the Energy-Channel relationship and the energy resolution are
calibrated in two working modes (Normal-Gain mode \& Low-Gain Mode) with the different detection energy ranges.
The simulative spectral analyses show that \emph{HXMT}/CsI can constrain the spectral parameters much better in the high energy band than that in the low energy band.
The joint spectral analyses are performed to ten bright GRBs observed simultaneously with \emph{HXMT}/CsI and other instruments (\emph{Fermi}/GBM, \emph{Swift}/BAT, Konus-\emph{Wind}), and the results show that the GRB flux given by \emph{HXMT}/CsI is systematically higher by $7.0\pm8.8\%$ than those
given by the other instruments. The \emph{HXMT}/CsI-\emph{Fermi}/GBM joint fittings also show that the high energy spectral
parameter
can be constrained much better as the \emph{HXMT}/CsI data are used in the joint fittings.

\textbf{Keywords:} instrumentation: detectors --- space vehicles: instrumentation --- telescopes --- gamma rays: general

\end{abstract}
\maketitle

\section{Introduction}
The \emph{Hard X-ray Modulation Telescope}, dubbed as \emph{Insight-HXMT}, was originally proposed in the 1990s and launched on June 15, 2017
(\citealp{Tipei2006Hard,Ti2008The,Zhang2014}; \citealp{Overview}).
\emph{Insight-HXMT} consists of three collimating telescopes: the High Energy X-ray Telescope \citep[HE,][]{HEofHXMT},
the Medium Energy X-ray Telescope \citep[ME,][]{MEofHXMT} and the Low Energy X-ray Telescope \citep[LE,][]{LEofHXMT}.
The CsI detectors of \emph{Insight-HXMT}/HE can be used for monitoring gamma-ray bursts (GRBs), MeV-pulsars, solar flares, terrestrial gamma-ray flashes and
other gamma-ray sources. It can also search the electromagnetic counterpart of important astronomic events \citep[e.g., GW170817,][]{Li2018}.
The accurate instrumental response is essential input for gamma-ray data analysis, which in turn can provide more clues to our understanding of the universe
in Multi-Messenger Era \citep{Duncan1992Formation,SariSpectra,Piran2004The,Abbott2017GW170817,Abbott2017Multi}. Hence, the calibration of the \emph{Insight-HXMT}/HE CsI detectors (\emph{HXMT}/CsI) is essential for the gamma-ray data analysis of \emph{Insight-HXMT}.

The instrumental response of \emph{HXMT}/CsI can be divided into two parts: the energy redistribution of the photons from
incident energy to deposition energy that determined by the property of the CsI crystal; or the Energy-Channel (E-C) relationship that determined
by the electronic system of the instrument.
For the photon energy redistribution,
the reliability is mainly determined by the accuracy of the mass model of the satellite and the payloads \citep{Agostinelli2003G}.
The initial mass model of the satellite platform \citep{Xie2015} is too simplistic to generate accurate instrumental response for the all sky gamma-ray monitoring.
Thus we calibrate the mass model with the Crab pulse radiation as a standard candle \citep{Li2018}.
The in-orbit E-C relationship and energy resolution of the instrument can be obtained by analyzing
the emission lines of the in-orbit observed and the on-ground simulated background spectra \citep{groundofHE}.
The E-C relationship varies over time, whereas the energy resolution remains stable.
Therefore, we update the E-C relationship every month, and take the average energy resolution of all calibration results to generate the instrumental response.
After the above calibration, a new response matrix library is established and a simulative spectral analysis is performed to test the \emph{HXMT}/CsI
spectral capabilities.

In a GRB observation, the incident direction of the GRB photons is supposed to be arbitrary, however, only the instrumental response to several directions can be calibrated directly.
A common method of the instrumental response testing is the cross-calibration with other instruments by comparing the energy spectrum of the simultaneously observed GRB
\citep{Sakamoto2011,Tsujimoto2011,Tierney2011The,Ishida2014}. The detection efficiency of \emph{HXMT}/CsI is checked by the joint spectral analyses with \emph{Fermi}/GBM,
\emph{Swift}/BAT and Konus-\emph{Wind}, in which we find that \emph{HXMT}/CsI can provide better constraint on GRB spectrum at higher energy band.

This study is organized as follows. In Section~\ref{sec:HXMT_CsI}, a description of \emph{Insight-HXMT}/HE CsI detectors is given.
In Section \ref{sec:calibration}, we show the calibration of the instrumental response, including the mass model, the E-C relationship and the energy resolution.
In Section~\ref{sec:Response matrix and simulated spectrum fitting}, the simulative spectral analyses with the calibrated response matrix library
are performed to show the spectral capability of \emph{HXMT}/CsI.
In Section \ref{sec:Joint-Fitting}, we present the results of the joint spectral analyses with other instruments.
Finally, the discussion and conclusion are given in Section~\ref{sec:Dis_Con}.

\section{\emph{Insight-HXMT}/HE CsI Detectors}
\label{sec:HXMT_CsI}
\begin{figure}[!htbp]
  \centering
  \includegraphics[width=0.5\textwidth]{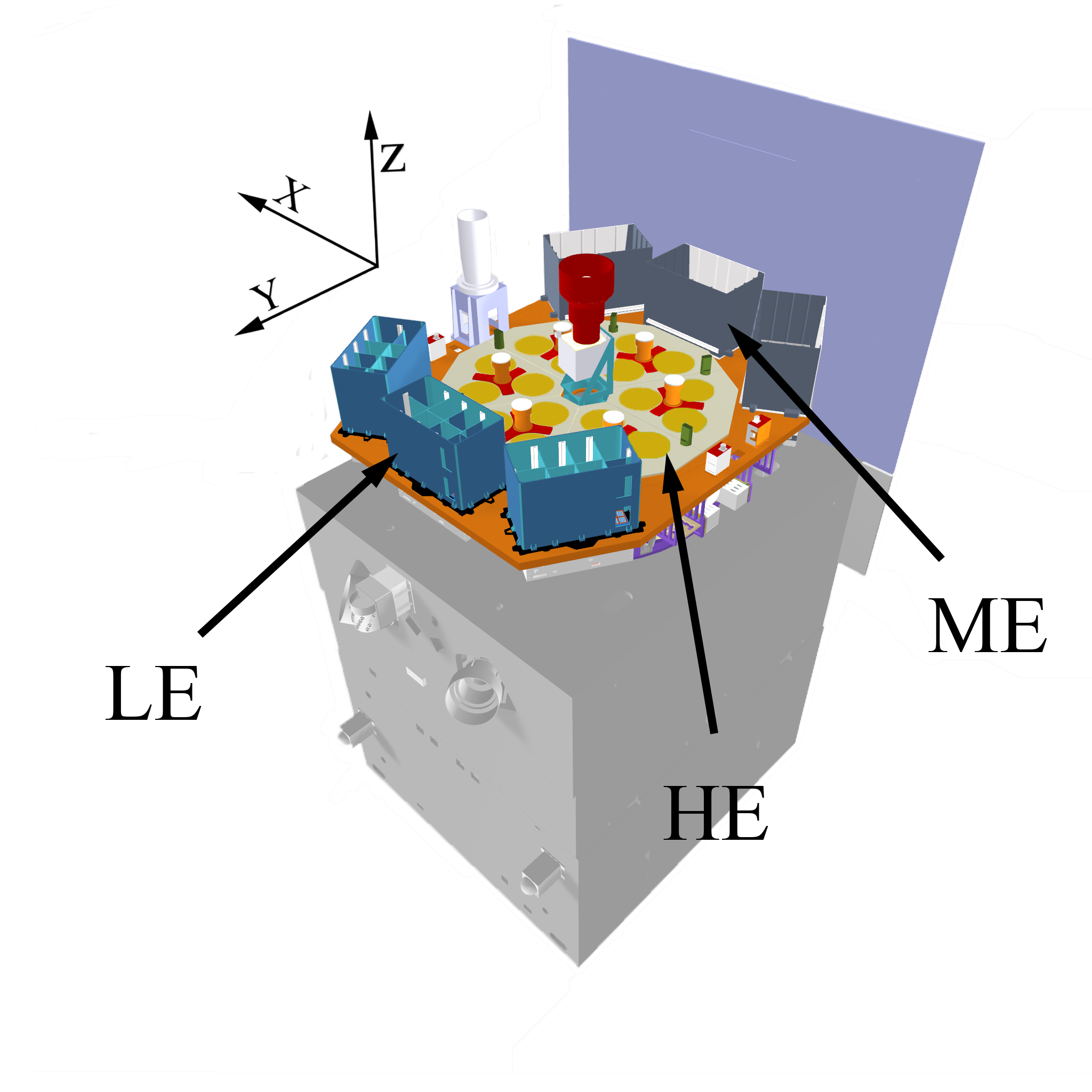}\\
  \caption{The illustration of the satellite platform and the payloads of \emph{Insight-HXMT}. The coordinate system used in this study is shown on the upper-left.
  }\label{HEjpg}
\end{figure}

\emph{Insight-HXMT}/HE is an X-ray space-born telescope, designed for carrying out the pointing observation, scanning observation,
and monitoring of GRBs (\citealp{Tipei2006Hard,Ti2008The,Zhang2014}; \citealp{Overview,HEofHXMT}). It consists of 18 NaI(Tl)/CsI(Na) phoswich detectors (HED-0, 1 ... to 17)
, each with a diameter of 190 mm and a collimator with the top coverage composed of plumbum and tantalum.
The NaI detectors are sensitive to the hard X-rays in $20-250$~keV, whereas the CsI acts as anticoincidence
detector to reduce upward background.
The thickness of the NaI(Tl) and CsI(Na) crystals are 3.5 mm and 40 mm, respectively \citep{HEofHXMT}.

Gamma-ray photons with energies $>200$~keV can penetrate the spacecraft and payload structure and leave their footprints in \emph{Insight-HXMT}/HE (Figure~\ref{HEjpg}).
Due to the limitations of the crystal thickness and the small field of view blocked by collimators, it is difficult to detect the GRBs by the NaI(Tl) detectors.
However, thanks to the high thickness of the CsI(Na) crystals, high energy gamma-ray photons can be recorded by the CsI detectors,
thus the GRBs not occulted can be detected.

The CsI detectors have two working modes in different detection energy ranges,
i.e., the Normal-Gain (NG) mode in $80-800$~keV and the Low-Gain (LG) mode in $200-3000$~keV (both refer to the deposited energies).
NG mode is the main working mode that the auto-gain control system can keep the full-energy-peak of the 59.5 keV photons (emitted from a radioactive source $^{241}$Am) in a fixed channel of the NaI detector \citep{HEofHXMT}. However, the auto-gain control system cannot keep the \emph{HXMT}/CsI E-C relationship stable. In LG mode, the high voltage of the each \emph{HXMT}/HE detector is reduced and the auto-gain control system is disabled to achieve a higher energy range detection. As derived from the two-year observations since \emph{Insight-HXMT} has operated in-orbit, the E-C relationships of each \emph{HXMT}/CsI detector for both NG and LG modes vary over time (the details will be described in Section \ref{sec:EC_ER}).

\begin{figure*}[t]
    \centering
    \subfigure{
    \begin{minipage}[t]{0.5\linewidth}
    \centering
    \includegraphics[width=0.8\linewidth]{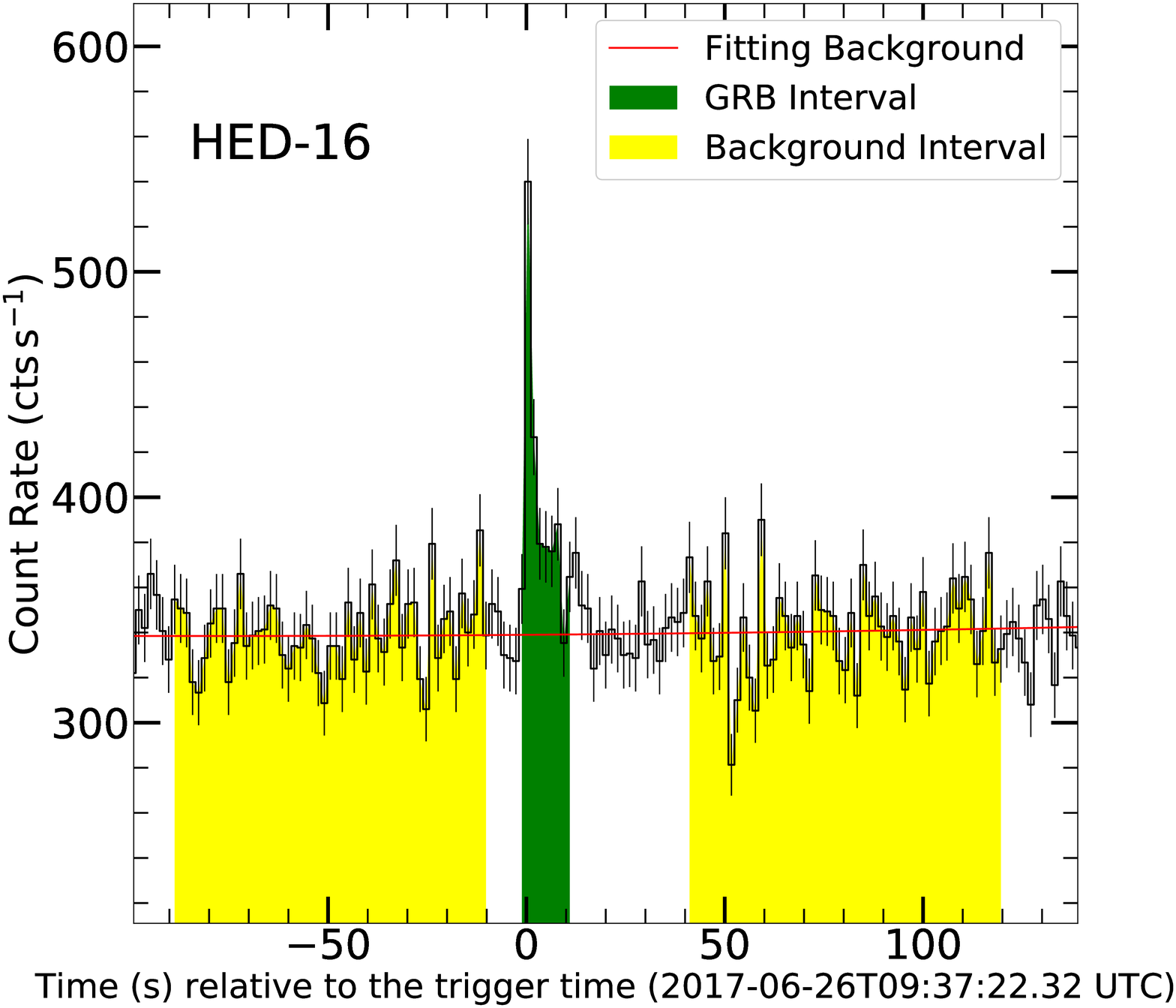}
    \end{minipage}%
    }%
    \subfigure{
    \begin{minipage}[t]{0.5\linewidth}
    \centering
    \includegraphics[width=0.8\linewidth]{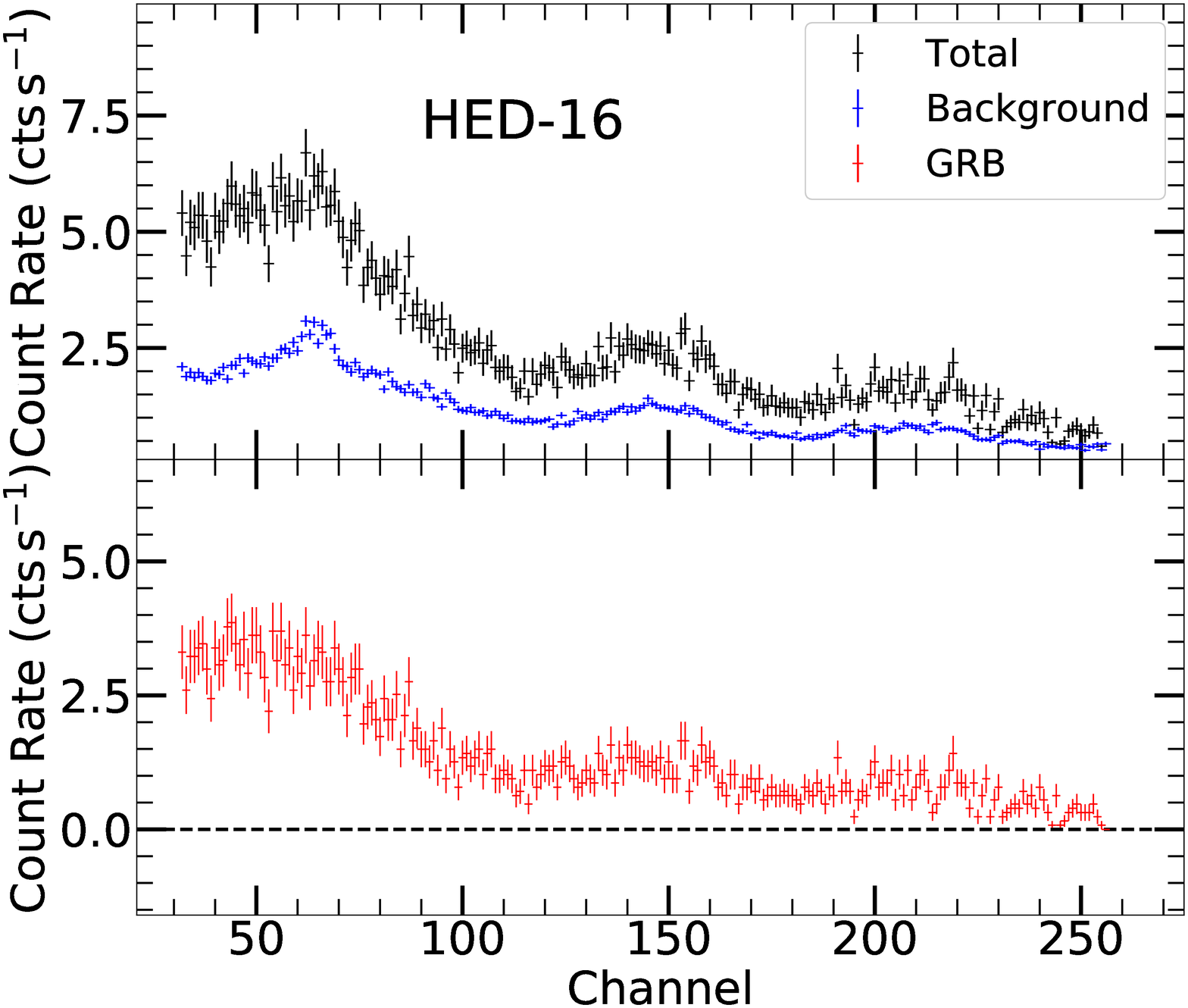}
    \end{minipage}%
    }%
    \centering
    \caption{Light curve (left) and spectrum (right) of HED-16 for GRB 170626A. In the left panel, the green region is the GRB interval, the yellow region are the two background intervals. In the right panel, the black, blue, red data are the total GRB spectrum, the background spectrum, and the net GRB spectrum, respectively.}\label{lc_spec}
\end{figure*}

Figure \ref{lc_spec} shows the observation of GRB 170626A by \emph{HXMT}/CsI (HED-16) in NG mode. In Figure \ref{lc_spec}, the left panel shows the light curve ($80-800$ keV) covering -100 s to 140 s with respect to the trigger time of the GRB. We first obtain the total spectrum in burst duration (-1.2 s to 11.7 s, green region in the left panel), which contain both GRB and background.
The \emph{HXMT}/CsI background is observed to have stable spectral shape on time scale of a few hundred seconds, and smoothly varying intensity that can be described with a quadratic function.
Thus the background can be estimated from the time intervals before and after the GRB trigger.
For GRB 170626A, the background intervals are selected as -90 s to -10 s and 40 s to 120 s with respect to the GRB trigger (yellow regions). The background intensity is fitted with a quadratic function, and then the background intensity ratio of GRB interval to elsewhere is calculated for correcting the GRB background spectrum (blue in the right panel). Finally, the background spectrum is subtracted off and the resulted net GRB spectrum (red in the right panel) can be used for spectral analysis. In the following, all the spectral analyses are performed the spectrum merged over the 18 \emph{HXMT}/CsI detectors, with the XSPEC tool with version 12.10.0c.

The coordinate system used in this study is shown in Figure \ref{HEjpg}.
The incident angle $\theta$ is the angle between source direction and the Z-axis, and
the azimuthal angle $\phi$ follows the right-hand rule where the X-axis is $\phi=0^\circ$.

\section{Calibration}
\label{sec:calibration}
The calibration of the instrumental response is mainly composed of two parts, i.e., the calibration of the mass model
and the calibration of the E-C relationship and energy resolution. The details will be described in follows.

\subsection{Mass Model}
\label{sec:MM}
Geant4 is a software package \citep{Agostinelli2003G} that is widely used in the space telescope projects,
e.g., \emph{XMM-Newton}, \emph{Swift} and \emph{Suzaku} \citep{Hall2008Simulating,HallSpace,Mineo2017An,FiorettiGeant4}.
With Geant4 tools and mass model of the instrument that describe the geometrical distributions and material compositions of the complex detectors,
Geant4 package can simulate the whole transport processes of the particles through matter, including the interactions between
all the known particles and the detector materials.

The instrumental response can be obtained from the Geant4 simulations given with the mass model and the incident photons of different energies.
Compared with the real instruments, the mass model used in the Geant4 simulation is simplified to different levels.
As described in \cite{Li2018}, the stable pulse emission of Crab at hard X-rays can be detected with \emph{HXMT}/CsI in various incident directions ($\theta$, $\phi$).
Thus the observed count rate of the Crab pulse component is used to test acceptance of the simplified mass model.
With the spectral parameters in \cite{Kuiper2001The} and the simulated response, the ratio of count rate between the observed and the simulated can be obtained for Crab pulse component.

\begin{figure*}[t]
\centering
    \tikzstyle{startstop} = [rectangle, rounded corners, minimum width = 2cm, minimum height=1.5cm,text centered, draw = black]
    \tikzstyle{decision} = [diamond, aspect = 3, text centered, draw=black]
    \tikzstyle{arrow} = [->,>=stealth]
    \begin{tikzpicture}[node distance=2cm]
    \node[startstop](start1){Mass Model};
    \node[startstop,right of = start1, xshift = 1cm](start2){Geant4};
    \node[startstop,below of = start1, xshift = 1.5cm,yshift = -1cm](pro1){Response Matrix};
    \node[decision, below of = pro1, yshift = -0.5cm](dec1){$S_{\rm u}$ $<$ 0.3 ?};
    \node[startstop, below of = dec1, yshift = -0.5cm](stop1){Calibration done};
    \coordinate (point1) at (-2cm, -5.5cm);
    \coordinate (point2) at (1.5cm, -1.25cm);
    \draw (start1) |- (point2);
    \draw (start2) |- (point2);
    \draw [arrow] (point2) -- node [right] {Simulation} (pro1);
    \draw [arrow] (pro1) -- (dec1);
    \draw [arrow] (pro1) -- node [right] {Crab calibration} (dec1);
    \draw (dec1) -- node [above] {NO} (point1);
    \draw [arrow] (point1) |- (start1);
    \draw [arrow] (dec1) -- node [right] {YES} (stop1);
    \end{tikzpicture}
    \caption{The flow chart of the calibration of the mass model.}
    \label{fig:pro}
\end{figure*}
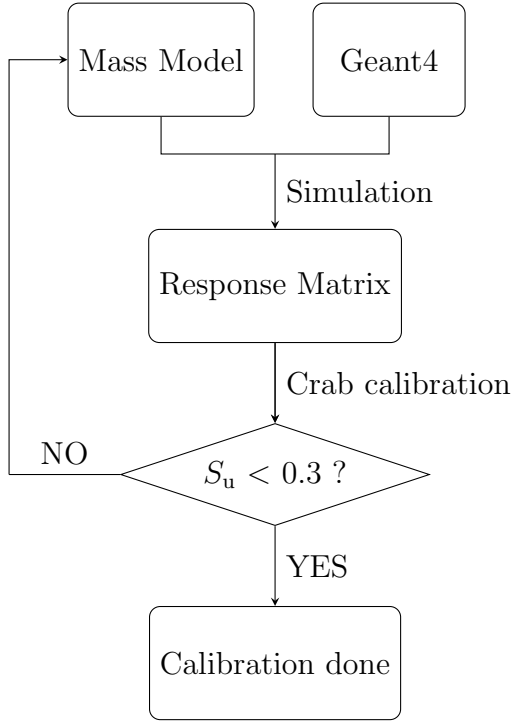

The intrinsic scatter of the ratio to unity is denoted as $S_{\rm u}$, and used as an indicator of the accuracy of the mass model, which is defined as:
\begin{equation}\label{func:Sin}
S_{\rm u}=\left\{\frac{1}{N-2}\sum _{i=1}^N \left[(r_i-1)^2 w_i-\delta _i^2\right]\right\}{}^{\frac{1}{2}},
\end{equation}
where
\begin{equation}\label{func:Sin}
w_i=\frac{\frac{1}{\delta _i^2}}{{\frac{1}{N}} \sum _{i=1}^N {\frac{1}{\delta _i^2}}},
\end{equation}
where $r_i$ refers the count rate ratio between the observed to the simulated, and $\delta _i$ the statistical errors of $r_i$.
With the process shown in Figure~\ref{fig:pro}, the mass model is updated iteratively until $S_{\rm u}<0.3$.
For the final mass model with $S_{\rm u}=0.26$, one has $r_i$ as shown in Table \ref{table0} for different incident directions.

\begin{table}[!htbp]
\caption{Ratios of the observed count rate to the expected count rate obtained with the final mass model.}
\begin{center}
\renewcommand\arraystretch{1.0}
\begin{tabular}{p{1.8cm}<{\centering}p{1.8cm}<{\centering}p{1.8cm}<{\centering}}
\hline
Observation number& Direction ($\theta$, $\phi$)& $r_i$\\
\hline
1& ($2^\circ$,$\;$$54^\circ$)& $1.01\pm0.25$\\
2& ($6^\circ$,$\;$$0^\circ$)& $0.82\pm0.09$\\
3& ($47^\circ$,$\;$$200^\circ$)& $0.87\pm0.33$\\
4& ($49^\circ$,$\;$$164^\circ$)& $1.10\pm0.29$\\
5& ($95^\circ$,$\;$$139^\circ$)& $0.69\pm0.27$\\
6& ($98^\circ$,$\;$$325^\circ$)& $0.77\pm0.15$\\
7& ($103^\circ$,$\;$$145^\circ$)& $1.03\pm0.25$\\
8& ($113^\circ$,$\;$$155^\circ$)& $0.71\pm0.12$\\
9& ($123^\circ$,$\;$$65^\circ$)& $0.86\pm0.16$\\
10& ($123^\circ$,$\;$$139^\circ$)& $0.92\pm0.13$\\
11& ($135^\circ$,$\;$$343^\circ$)& $0.73\pm0.06$\\
12& ($138^\circ$,$\;$$126^\circ$)& $0.63\pm0.09$\\
13& ($142^\circ$,$\;$$140^\circ$)& $1.09\pm0.26$\\
14& ($152^\circ$,$\;$$358^\circ$)& $0.48\pm0.27$\\
15& ($167^\circ$,$\;$$134^\circ$)& $0.44\pm0.18$\\
\hline
\end{tabular}
\end{center} \label{table0}
\end{table}

\subsection{Energy-Channel Relationship and Energy Resolution}
\label{sec:EC_ER}
\begin{figure*}[!htbp]
    \centering
    \includegraphics[width=4.4in]{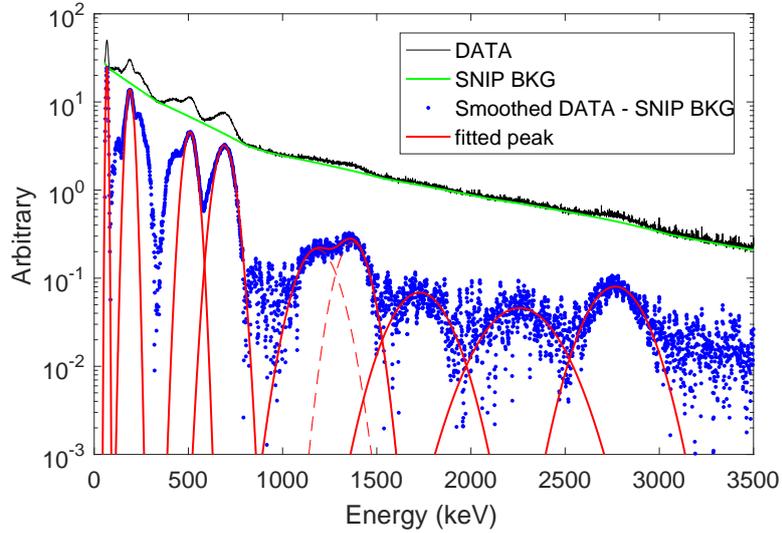}
    \caption{Geant4 simulation illustrating how to obtain the central energies of the bumps in the spectrum of the blank sky observation.
	The black data is the simulated spectrum by the Geant4 tools with the mass model and the on-ground measured energy resolution;
	the green line is the continuum obtained by the SNIP algorithm;
	the blue data is the background subtracted spectrum with only the bumps consist of several emission lines with the similar line energy.
	The red lines are the Gaussian functions that are used to fit the bumps to obtain the central energies of the bumps.}\label{CsI_spectrumjpg}
\end{figure*}

\begin{figure*}[htbp]
    \centering
    \subfigure{
    \begin{minipage}[t]{1.0\linewidth}
    \centering
    \includegraphics[width=1.00\textwidth]{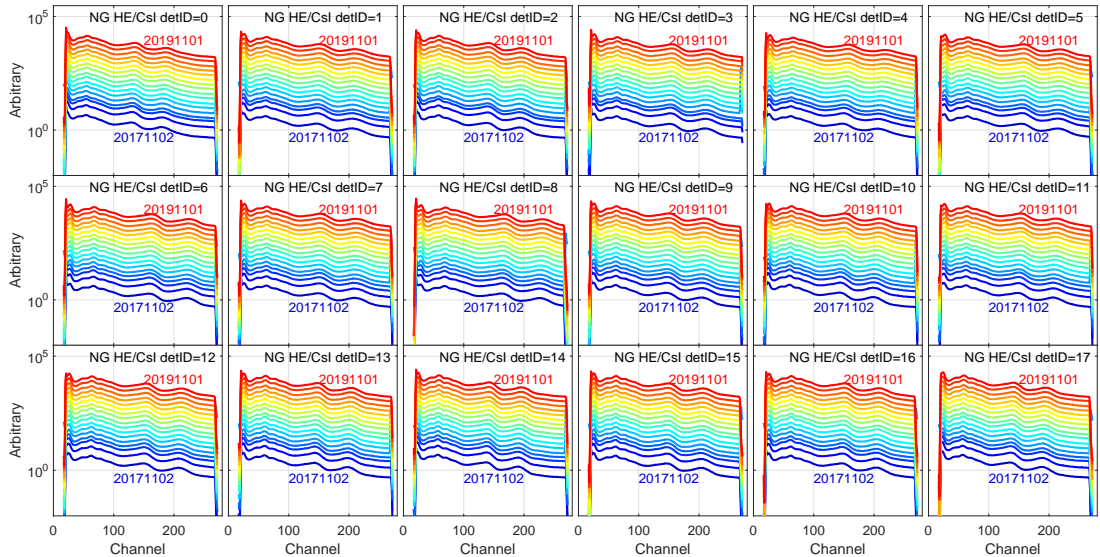}
    \end{minipage}\label{CsI_detected_spectrumjpg}%
    }%
	\caption{The observed spectra of all the 18 CsI detectors in NG mode among the two years as \emph{Insight-HXMT} operating in orbit, each line covers one or two months.
	The spectral amplitudes are adjusted to show the shifts of the spectra clearly.}\label{CsI_NG_detectedjpg}
\end{figure*}

\begin{figure*}[htbp]
    \subfigure{
    \begin{minipage}[t]{1.0\linewidth}
    \centering
    \includegraphics[width=1.00\textwidth]{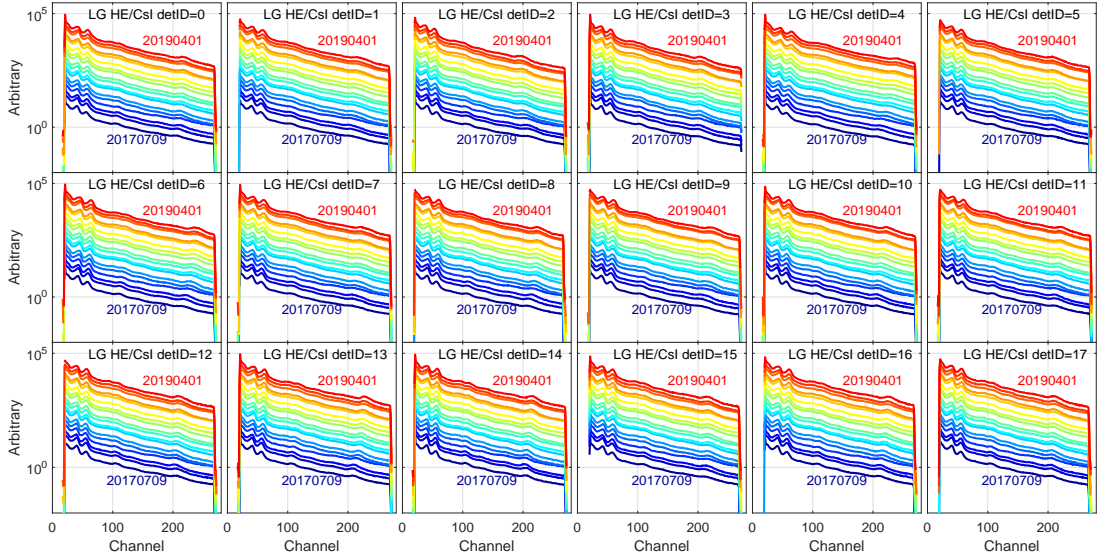}
    \end{minipage}\label{CsI_detected2_spectrumjpg}%
    }%
    \centering
    \caption{The same to Figure~\ref{CsI_NG_detectedjpg}, but for LG mode.}\label{CsI_LG_detectedjpg}
\end{figure*}

\begin{figure*}[t]
    \centering
    \subfigure{
    \begin{minipage}[t]{0.5\linewidth}
    \centering
    \includegraphics[width=1\linewidth]{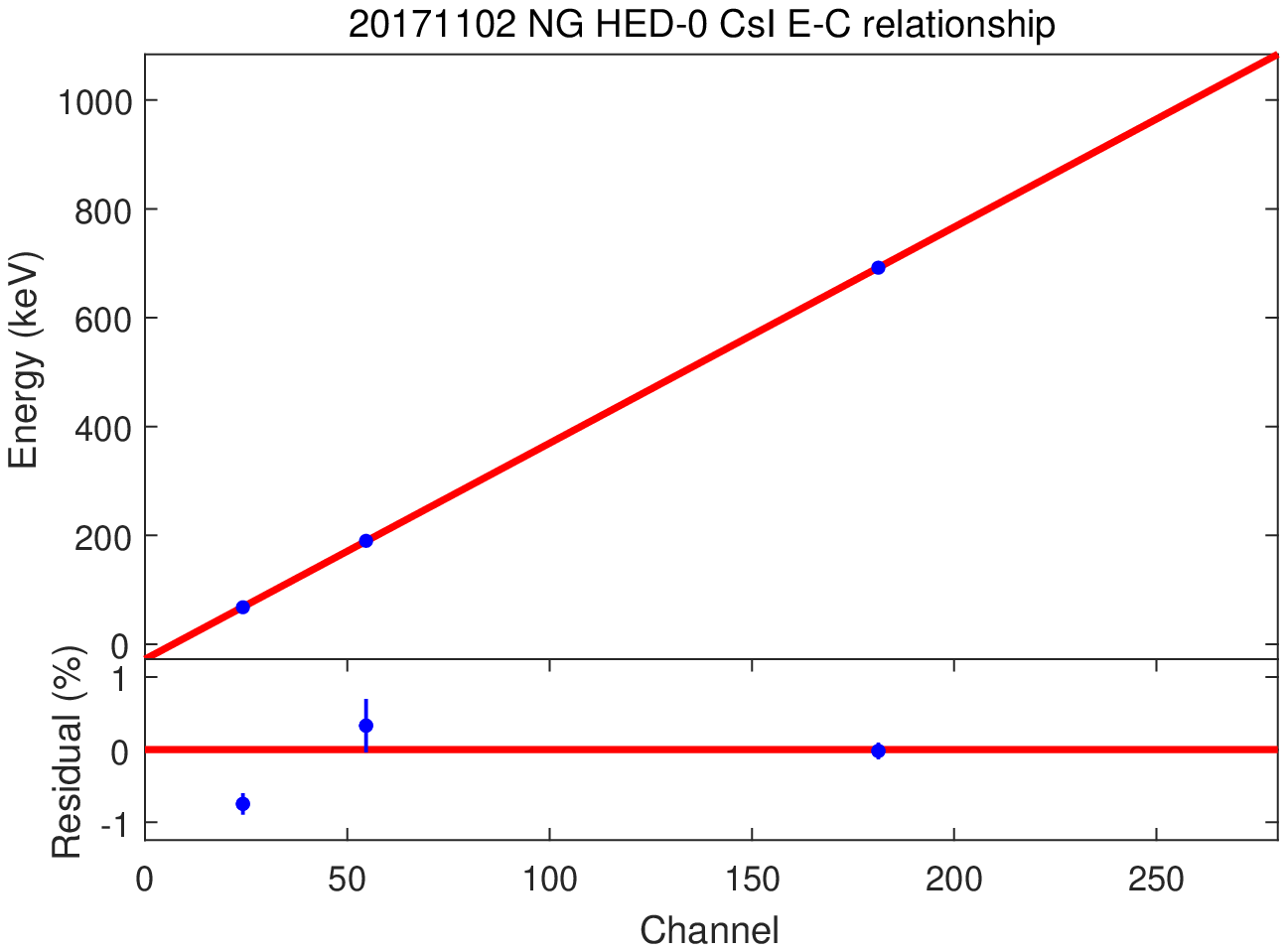}
    \end{minipage}%
    }%
    \subfigure{
    \begin{minipage}[t]{0.5\linewidth}
    \centering
    \includegraphics[width=1\linewidth]{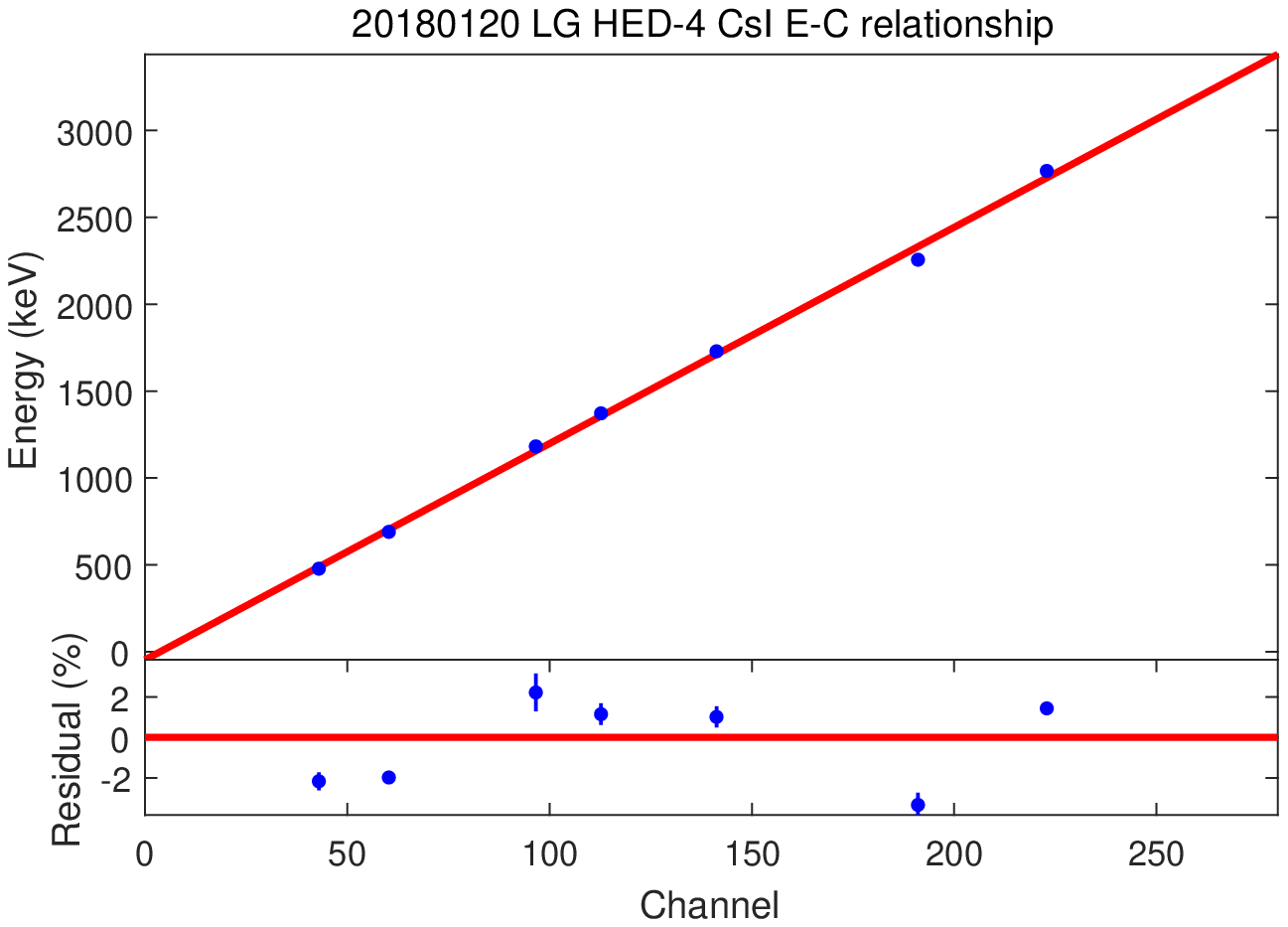}
    \end{minipage}%
    }%
    \centering
    \caption{Fittings of the E-C relationship of the CsI detectors in NG (HED-0, left) and LG (HED-4, right) mode.}\label{ec_fitting}
\end{figure*}

\begin{figure*}[t]
    \centering
    \subfigure{
    \begin{minipage}[t]{0.5\linewidth}
    \centering
    \includegraphics[width=3.2in]{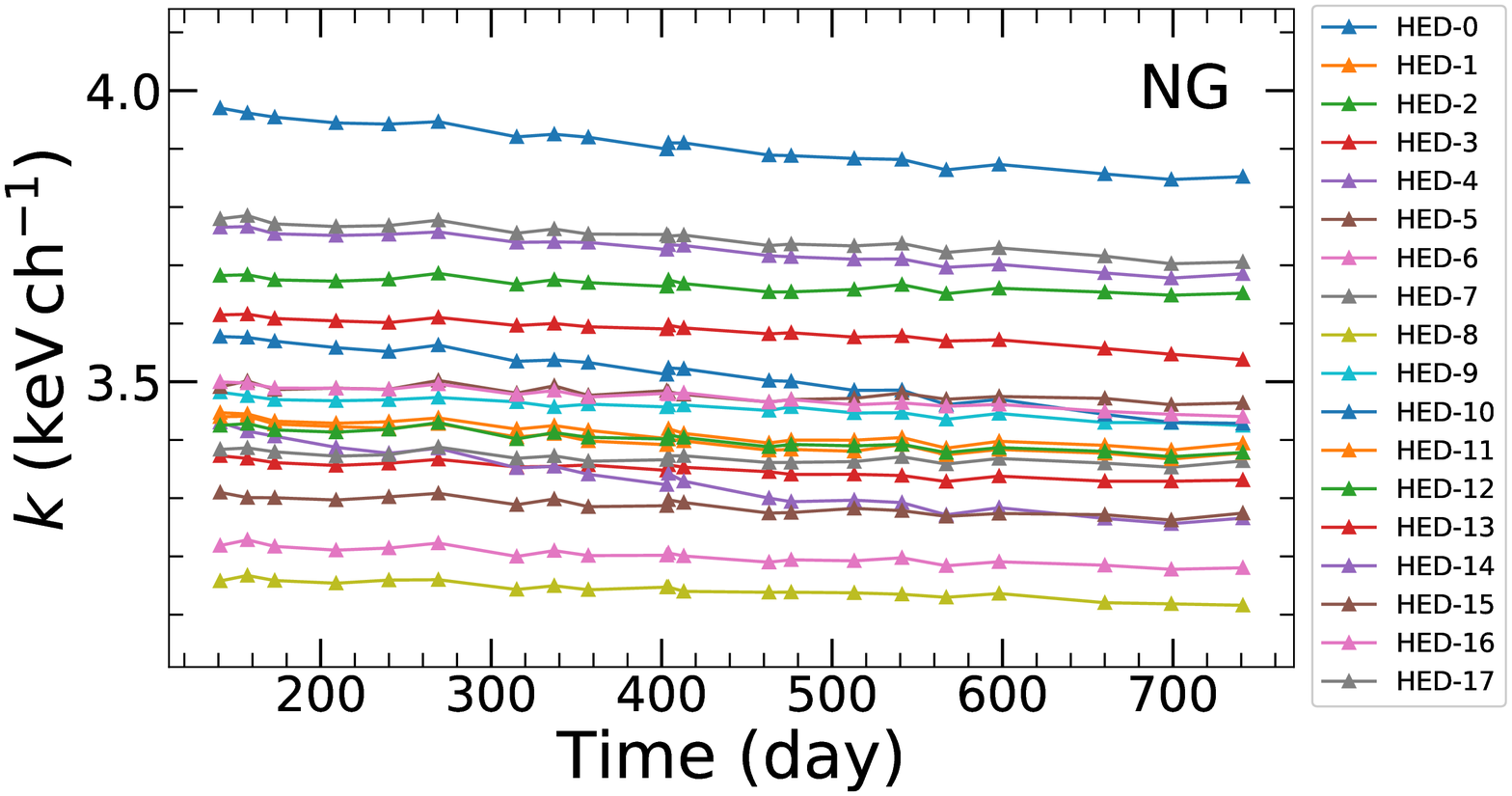}
    \end{minipage}\label{k_NG}%
    }%
    \subfigure{
    \begin{minipage}[t]{0.5\linewidth}
    \centering
    \includegraphics[width=3.2in]{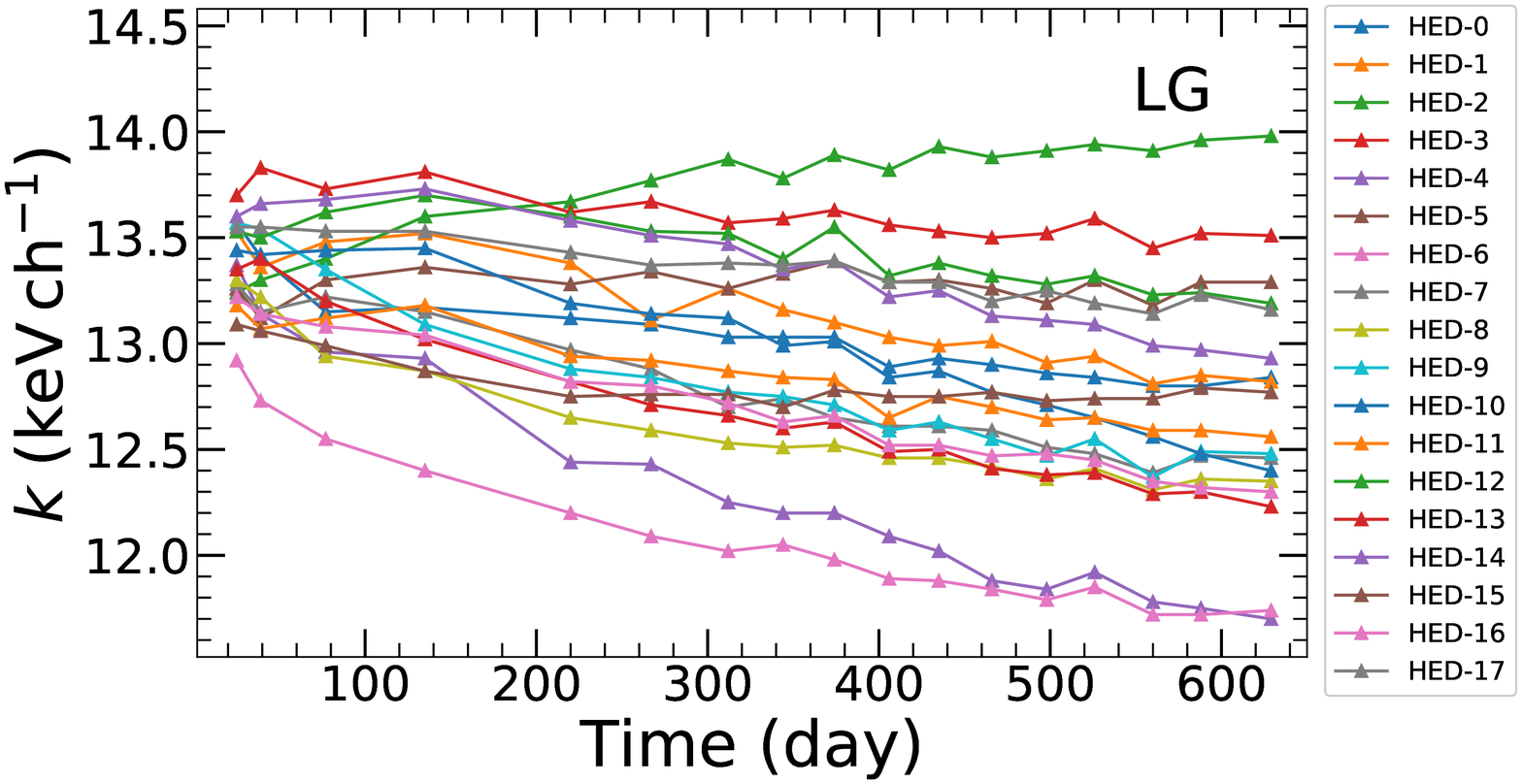}
    \end{minipage}\label{k_LG}%
    }%
    \centering
    \caption{Evolution of the coefficients $k$ in Equation~\ref{eq:EC} of the 18 CsI detectors in NG (left) and LG (right) mode.}\label{kjpg}
\end{figure*}

As the satellite operates in orbit, the E-C relationship and the energy resolution vary with decreasing fluoresce efficiency of the CsI crystals
and aging of the PMTs, thus they must be calibrated periodically. Here the blank sky observations and the data within earth shadow are adopted for calibrations in NG and LG modes, respectively. It is obvious that in both modes the observed spectra are rather structured, characterized with a series of bumps, each denoting combination of the emission lines residing in the background spectrum.
With the Geant4 tools and the mass model of \emph{Insight-HXMT}, as well as the energy resolution measured on ground, the simulation can result in a spectrum very similar to the observed one.
For further investigation of these bumps, first the continuum is measured with the Statistics-sensitive Nonlinear Iterative Peak-clipping (SNIP) algorithm
\citep{Morh1997Background,Morh2007Multidimensional} and then subtracted off.
As shown in Figure \ref{CsI_spectrumjpg}, the bump structures stand out clearly in the residuals. To the first step, each bump in both simulated (Figure \ref{CsI_spectrumjpg}) and observed spectra (Figure \ref{CsI_NG_detectedjpg} and Figure \ref{CsI_LG_detectedjpg}) is fitted with
a Gaussian function to represent the averaged energy and channel. Accordingly,
the E-C relationship can be obtained as,
\begin{equation}
	E(C) = kC + h,
\label{eq:EC}
\end{equation}
where $E$ is the deposited energy, $C$ the channel, $k$ and $h$ the fitting parameters.
As shown in Figure~\ref{ec_fitting}, deviation from such a linear relationship is less than $2\%$ for NG mode and $4\%$ for LG mode.
Since such a linear relation is evolving with time (Figure \ref{kjpg}), which is more significant in LG mode,
the E-C relationship shall be updated periodically. The current strategy is to calibrate the E-C relationship once per month for both modes.

\begin{figure*}[t]
    \centering
    \subfigure{
    \begin{minipage}[t]{0.5\linewidth}
    \centering
    \includegraphics[width=1\linewidth]{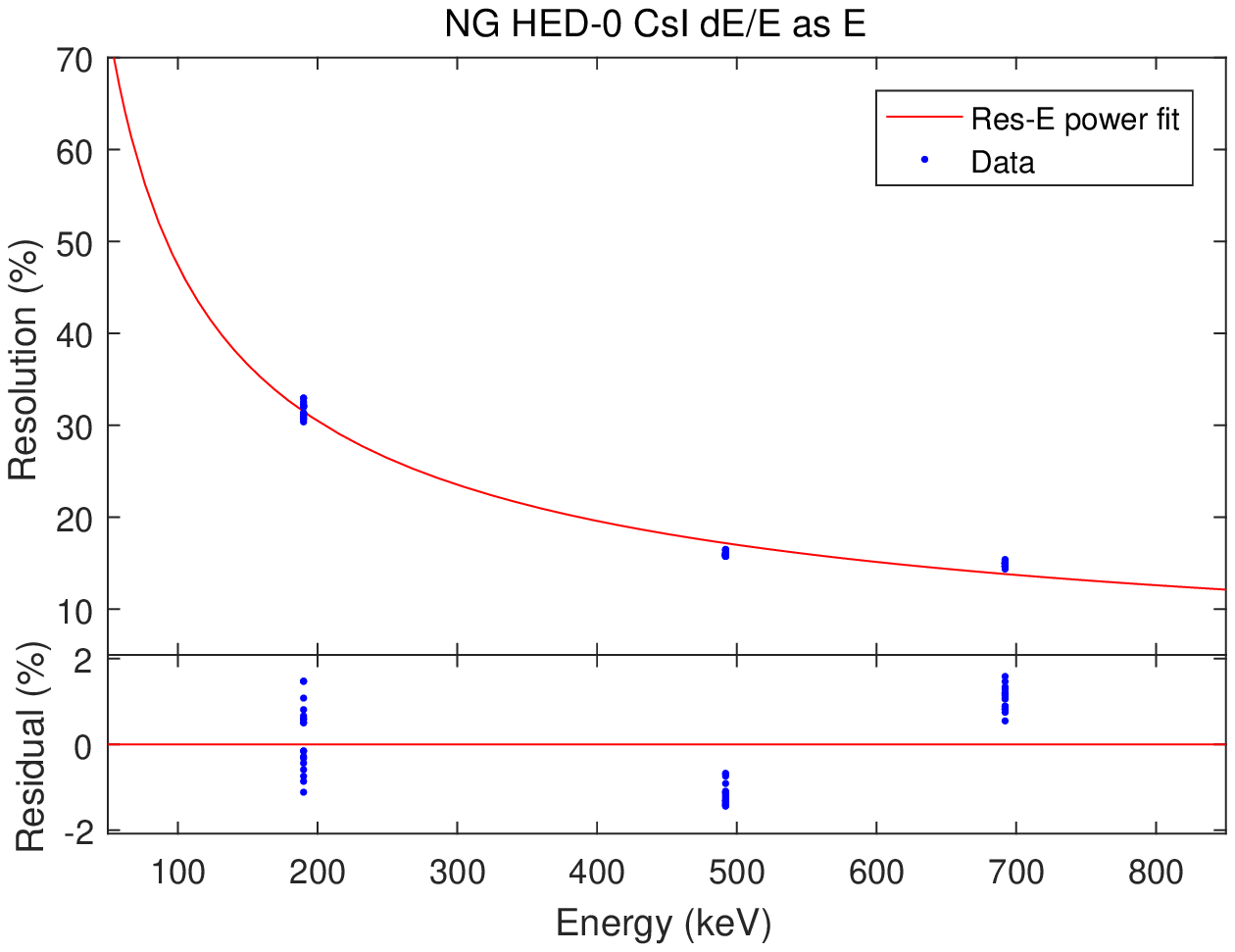}
    \end{minipage}%
    }%
    \subfigure{
    \begin{minipage}[t]{0.5\linewidth}
    \centering
    \includegraphics[width=1\linewidth]{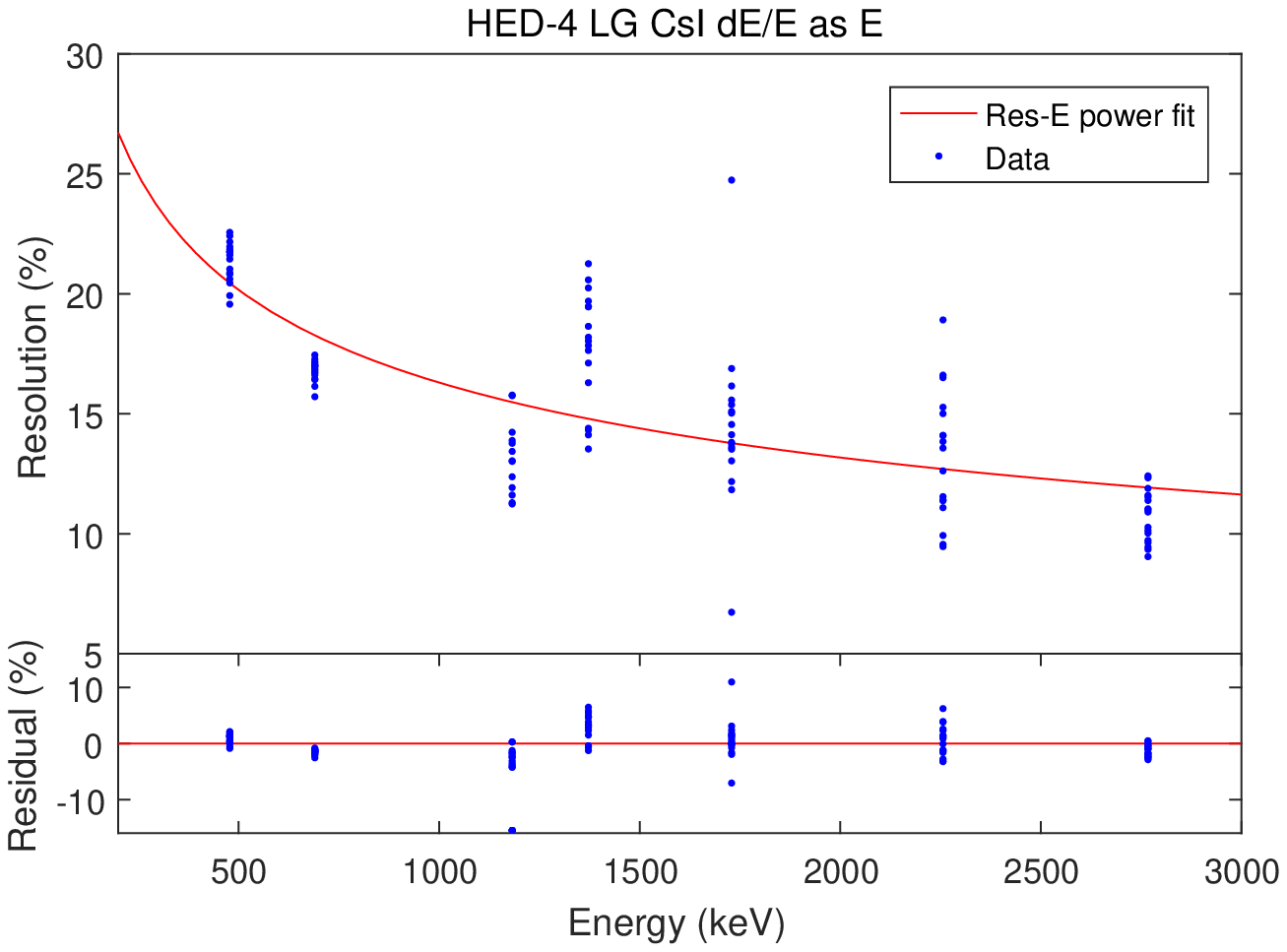}
    \end{minipage}%
    }%
    \centering
    \caption{Fittings of the energy resolution of the CsI detectors in NG (HED-0, left) and LG (HED-4, right) mode.}\label{resolution}
\end{figure*}

\begin{figure*}[t]
    \centering
    \subfigure{
    \begin{minipage}[t]{0.5\linewidth}
    \centering
    \includegraphics[width=3.2in]{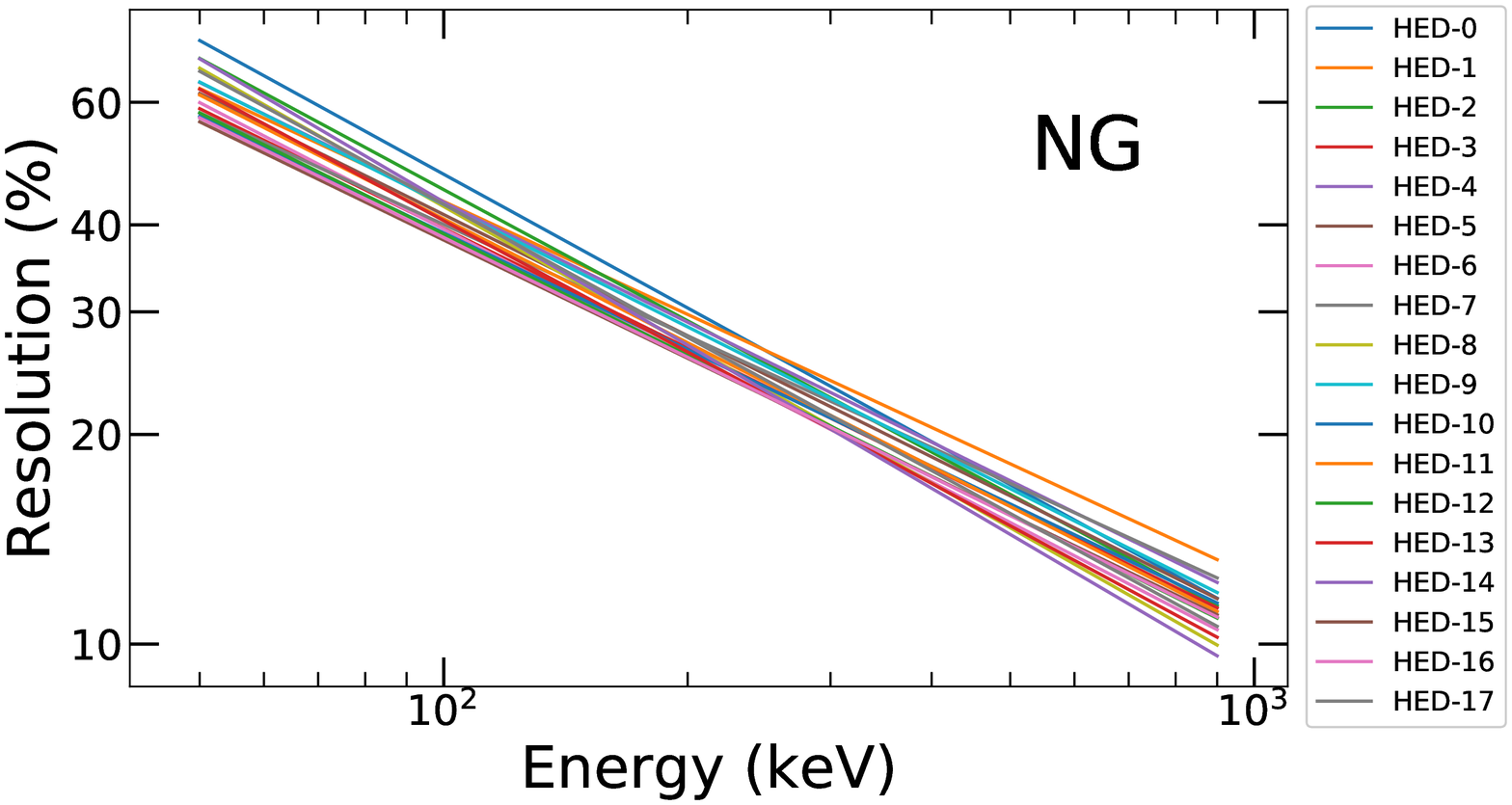}
    \end{minipage}\label{r_NG}%
    }%
    \subfigure{
    \begin{minipage}[t]{0.5\linewidth}
    \centering
    \includegraphics[width=3.2in]{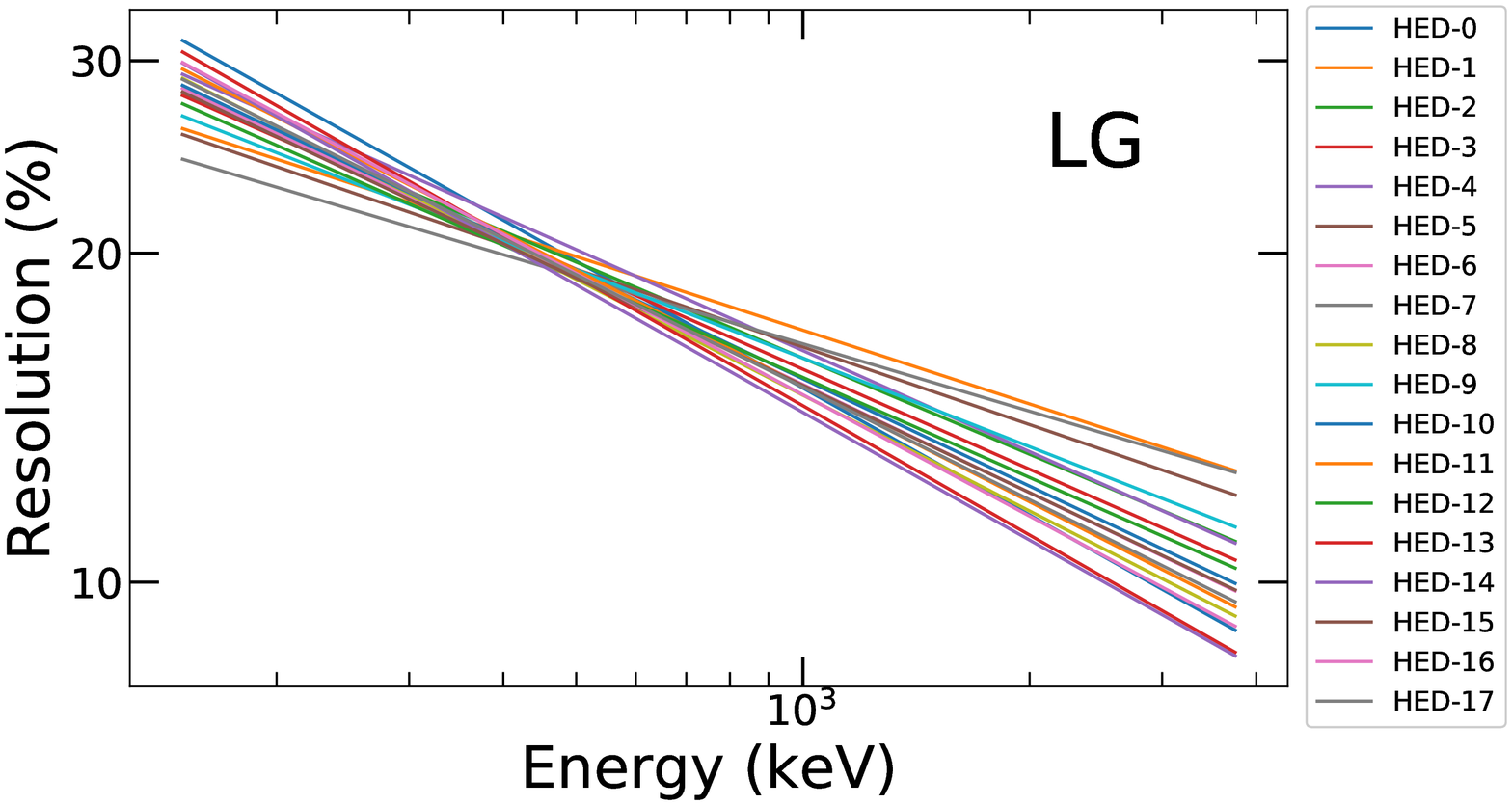}
    \end{minipage}\label{r_LG}%
    }%
    \centering
\caption{Energy resolutions of the 18 CsI detectors in NG (left) and LG (right) mode.}\label{rjpg}
\end{figure*}

After the E-C relationship is determined, the energy resolution can be obtained by fitting each bump in the observed spectrum with
the emission lines modeled with Geant4 simulation. Each bump can be fitted by several emission lines with width characterized by
the energy resolution at the average energy of the bump.
The relationship of the energy resolution versus energy is fitted by an exponential function (Figure~\ref{resolution}),
\begin{equation}
	R(E) = aE^b,
\label{eq:ER}
\end{equation}
where $R(E)$ is the energy resolution at the deposited energy $E$; $a$ and $b$ are the fitting parameters, respectively.
Based on the monthly monitoring of each bump width, we find that
the energy resolutions are basically stable over time.
Therefore, we
obtain the time-averaged $a$ and $b$ by fitting simultaneously all the monthly-segmented data with Equation~\ref{eq:ER}.
The deviation with such a fit for bumps is in general
within $2\%$ in NG mode and within $10\%$ in LG mode.
Examples are shown Figure~\ref{resolution} for HED-0 in
NG mode and HED-4 in LG mode.
Finally, the energy resolution is derived which, as shown in Figure~\ref{rjpg}, shows slightly dispersion among different detectors.

\section{Response Matrix and Spectral Analysis via Simulation}
\label{sec:Response matrix and simulated spectrum fitting}
\begin{figure*}[t]
    \centering
    \subfigure{
    \begin{minipage}[t]{0.33\linewidth}
    \centering
    \includegraphics[width=2.4in]{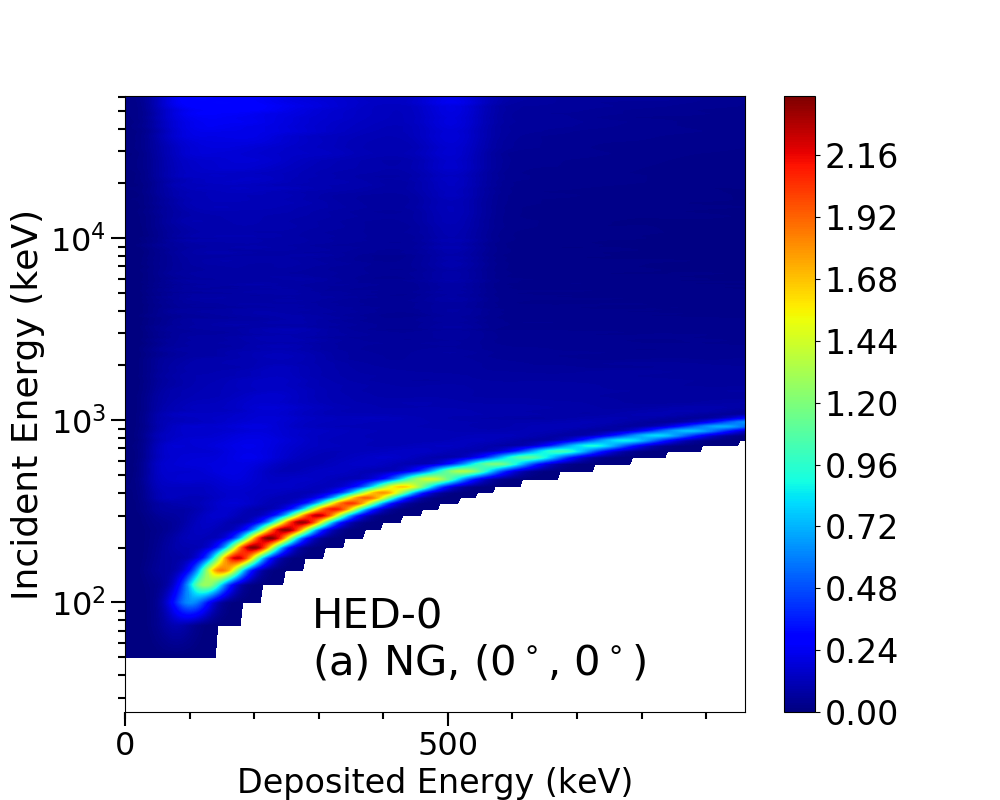}
    \end{minipage}%
    }%
    \subfigure{
    \begin{minipage}[t]{0.33\linewidth}
    \centering
    \includegraphics[width=2.4in]{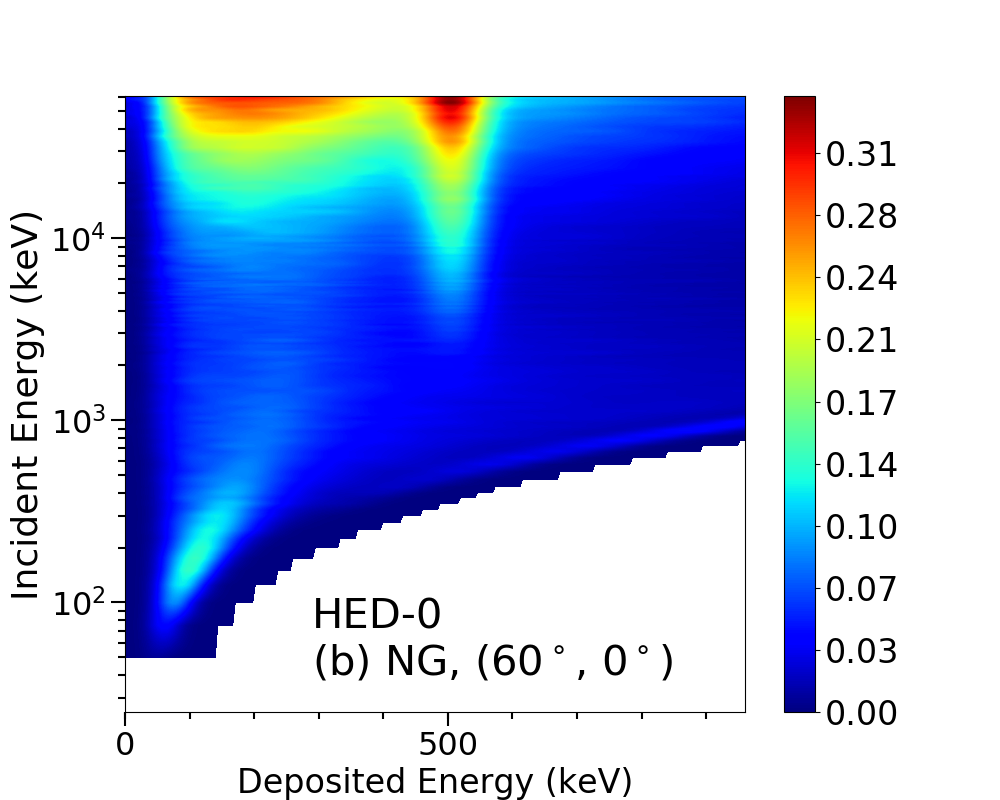}
    \end{minipage}%
    }%
    \subfigure{
    \begin{minipage}[t]{0.33\linewidth}
    \centering
    \includegraphics[width=2.4in]{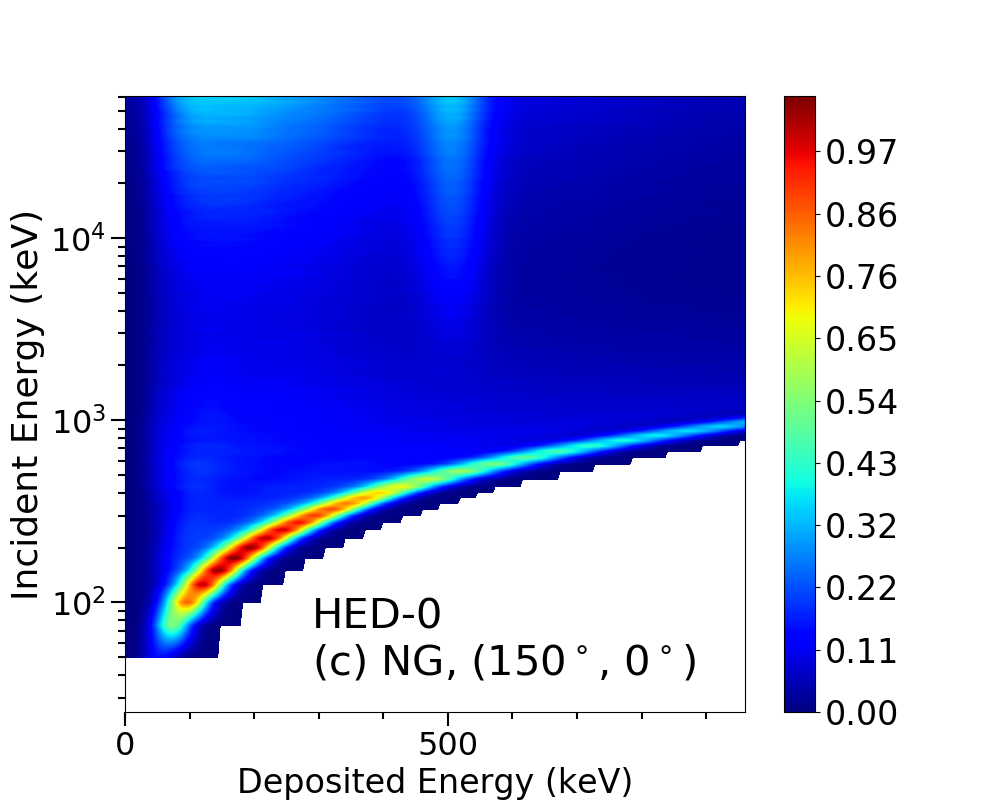}
    \end{minipage}%
    }%
    \quad
    \subfigure{
    \begin{minipage}[t]{0.33\linewidth}
    \centering
    \includegraphics[width=2.4in]{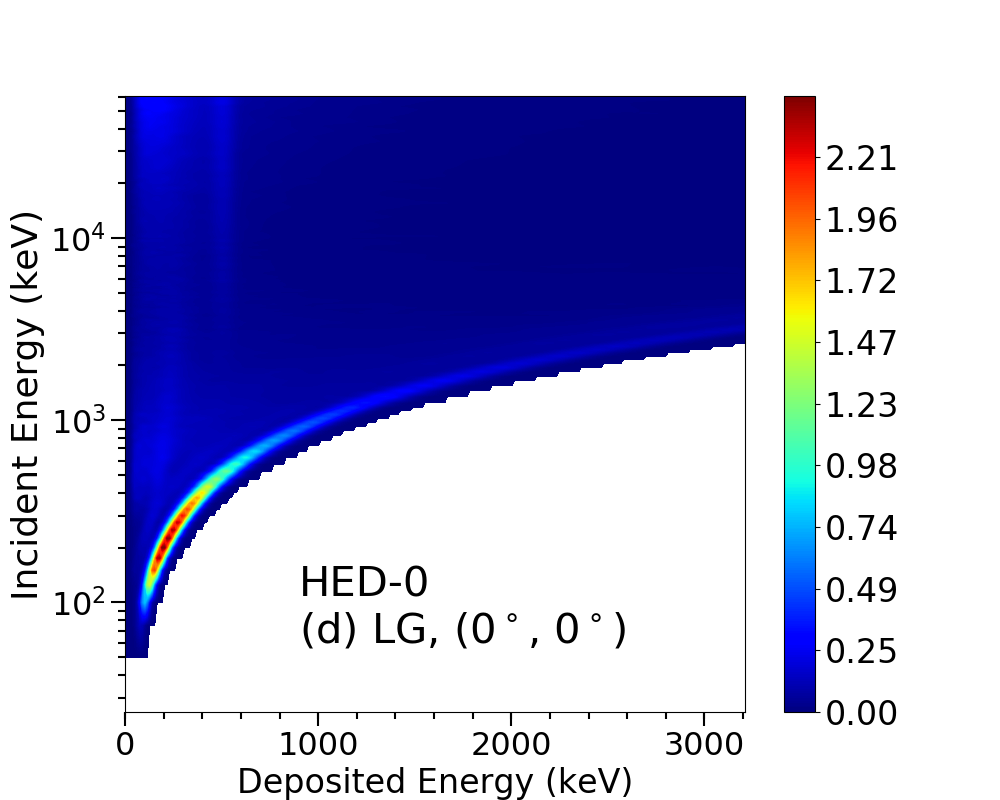}
    \end{minipage}%
    }%
    \subfigure{
    \begin{minipage}[t]{0.33\linewidth}
    \centering
    \includegraphics[width=2.4in]{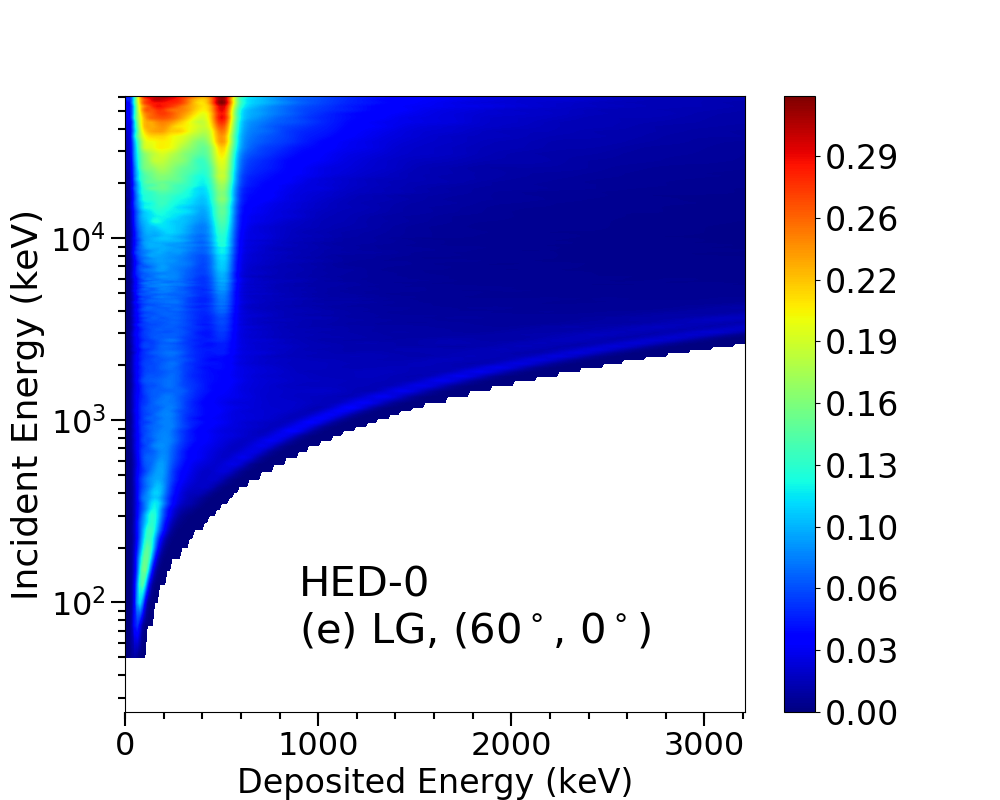}
    \end{minipage}%
    }%
    \subfigure{
    \begin{minipage}[t]{0.33\linewidth}
    \centering
    \includegraphics[width=2.4in]{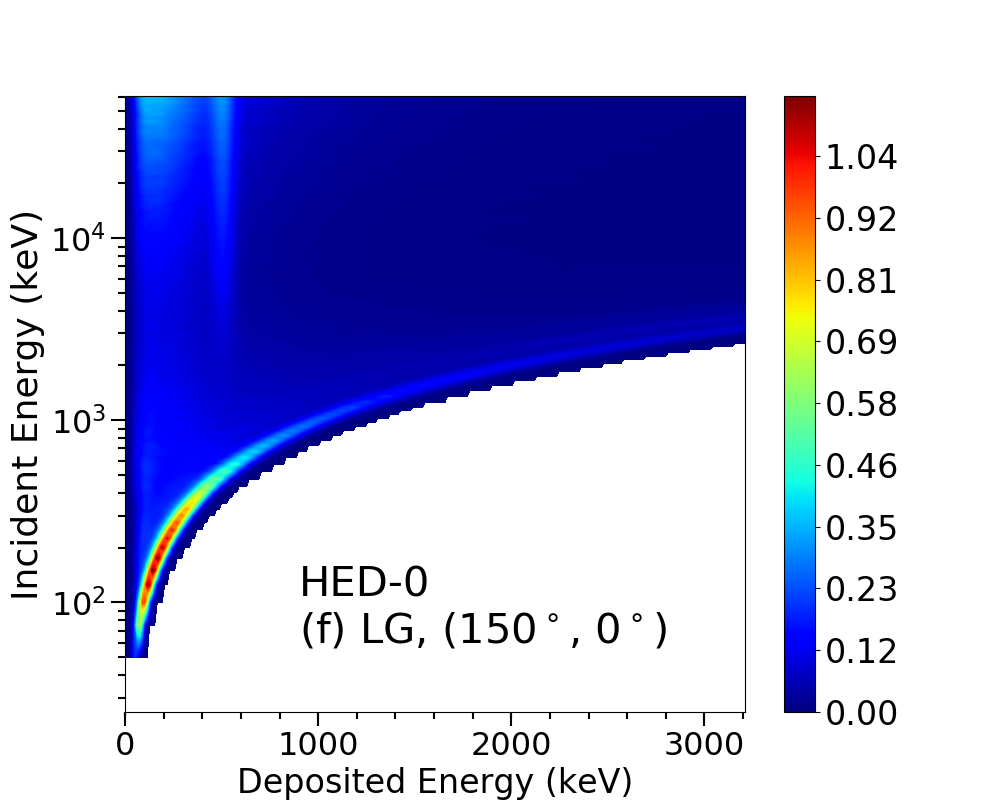}
    \end{minipage}%
    }%
    \centering
    \caption{Response matrices of the \emph{HXMT}/CsI detector (HED-0) with three incident directions in NG and LG mode. The X-axis and Y-axis are the deposition and incident energy, respectively.
	The corresponding unit of the color bar is cm$^2$.}\label{RSPjpg}
\end{figure*}

\begin{figure*}[t]
    \centering
    \subfigure{
    \begin{minipage}[t]{0.5\linewidth}
    \centering
    \includegraphics[width=2.5in]{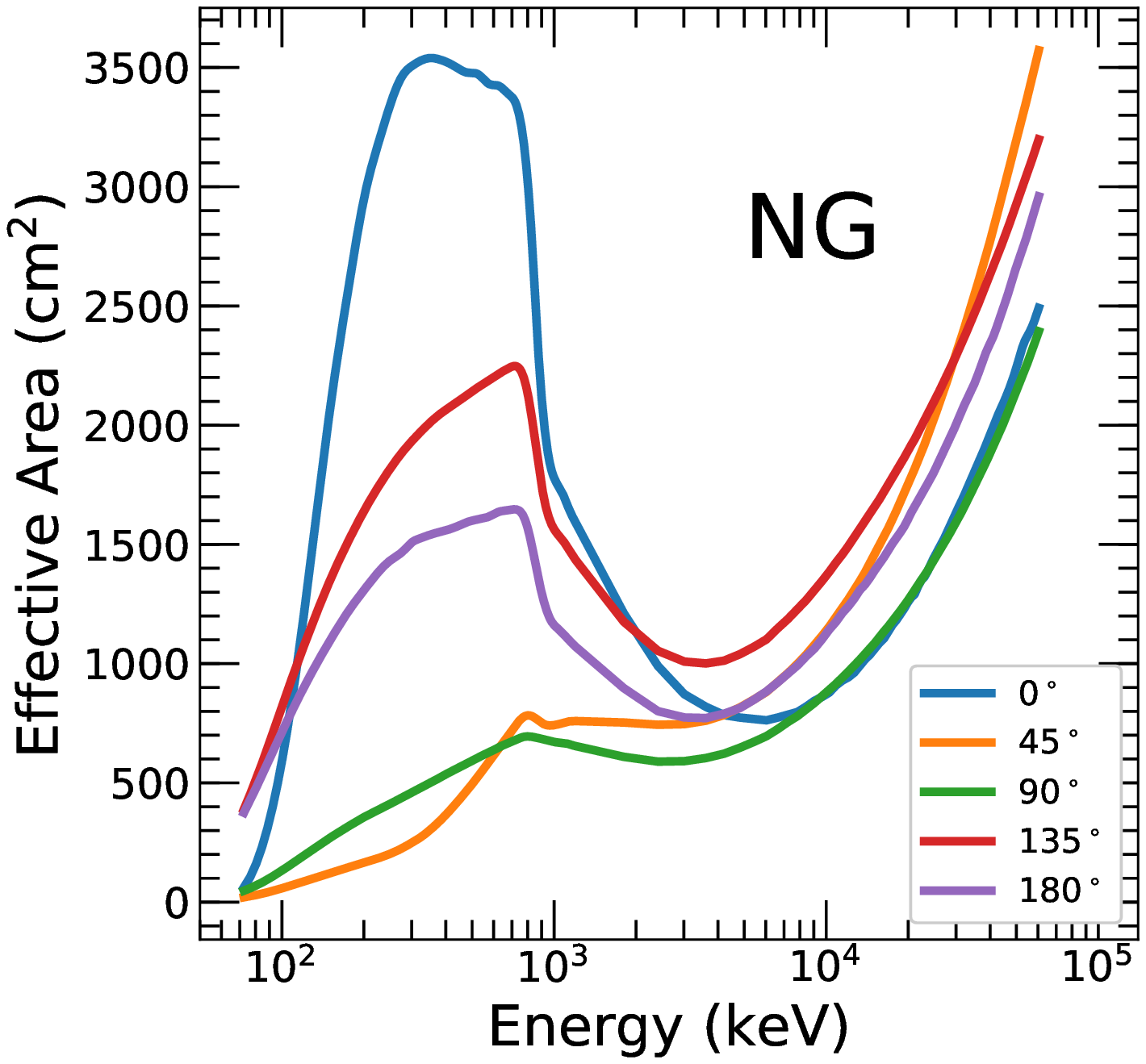}
    \end{minipage}%
    }%
    \subfigure{
    \begin{minipage}[t]{0.5\linewidth}
    \centering
    \includegraphics[width=2.5in]{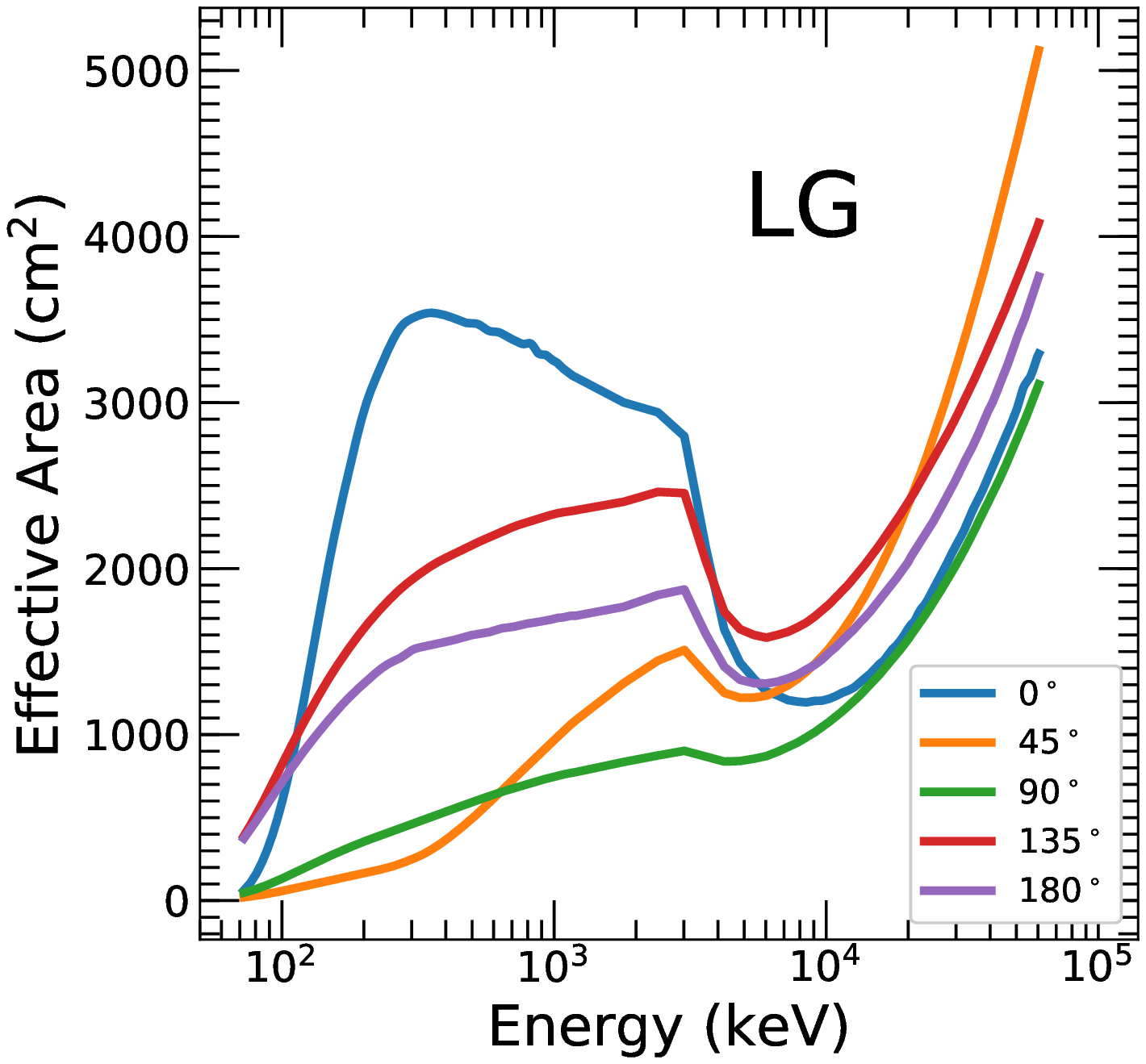}
    \end{minipage}%
    }%
    \centering
    \caption{Total effective area of 18 CsI detectors in NG (left) and LG (right) mode.
	Each line represents the effective area for each incident angle $\theta$ averaged in azimuthal angle $\phi$ from $0^\circ$ to $360^\circ$.}\label{ARFjpg}
\end{figure*}

The response matrix library of \emph{HXMT}/CsI can be obtained from Geant4 simulation in various incident directions with the final mass model.
Figure \ref{RSPjpg} shows the derived response matrices at three incident angles for both NG and LG modes. As shown in Figure \ref{ARFjpg}, the profiles of the effective areas are similar in NG mode and LG mode, and all the effective areas can be roughly divided into four stages:
\begin{enumerate}[(1)]
\item Rapid rise in $80-300$ keV: the absorption efficiency of CsI(Na) for photons is nearly $100\%$,
and the increase in effective area is dominated by the photon penetration effect.
Photons need to penetrate the Be window and NaI(TI) crystal to reach the CsI(Na) crystal at $\theta=0^\circ$,
and other satellite structures such as the collimators and the shielding rings
with high absorbtion efficiencies for photons at other incident angles.
The penetration efficiency increases with photon energy,
thus the effective area increases rapidly.
\item Slow decline ($\theta=0^\circ$) or slow rise ($\theta=45^\circ$, $90^\circ$, $135^\circ$, $180^\circ$) in $300-800$ keV for NG mode and $300-3000$ keV for LG mode:
the absorption efficiency of CsI(Na) gradually decreases with energy,
whereas the penetration efficiency to satellite structure increases with energy.
The effective area slowly decreases as the absorption effect is dominant ($\theta=0^\circ$), and slowly rises as the penetration effect is dominant (other incident angles).
\item Rapid decline in $800-900$ keV for NG mode and $3000-4000$ keV for LG mode:
the deposited energies of the gamma-ray photons beyond the upper threshold of the electronic system,
thus the signals cannot be recorded and the effective area drops sharply at the upper energy threshold.
\item Slowly decline to a minimum value and then rise in $>900$ keV for NG mode and $>4000$ keV for LG mode:
the CsI detectors can record the events caused by the interactions between the photons and satellite structures. The effect described in stage (3) is dominant before the minimum of the effective area, and then the effect of Compton scattering is dominant.
\end{enumerate}

\begin{figure*}[t]
    \centering
    \includegraphics[width=0.8\textwidth]{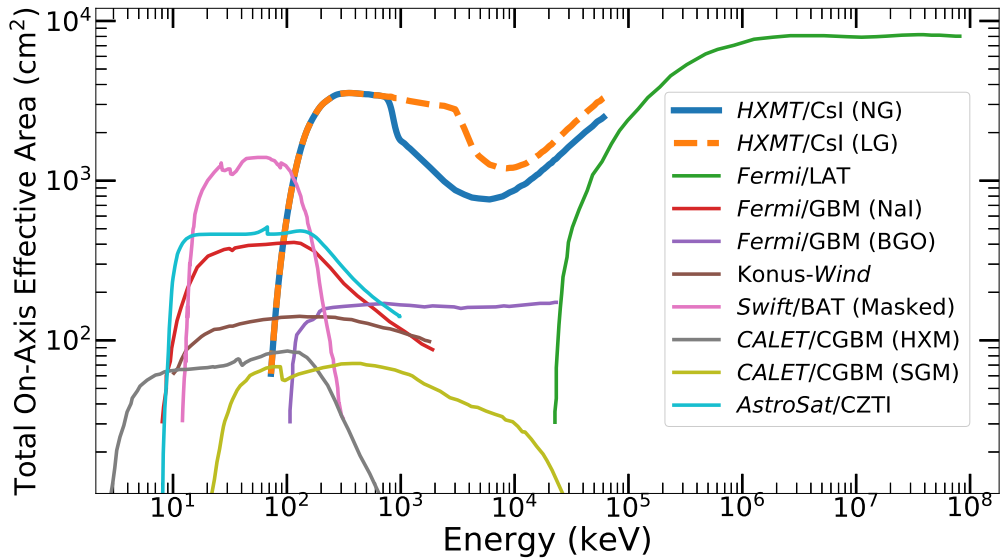}
    \centering
    \caption{Effective areas of \emph{HXMT}/CsI, \emph{Fermi}/LAT, \emph{Fermi}/GBM, Konus-\emph{Wind}, \emph{Swift}/BAT, \emph{CALET}/CGBM and \emph{AstroSat}/CZTI. The effective area of \emph{Fermi}/GBM (NaI) is the averaged over the unocculted sky.}\label{ARF_allsatjpg}
\end{figure*}

Due to the shielding effect of the collimators and
the shielding rings around the NaI/CsI crystals, the effective area at $6^\circ < \theta <90^\circ$ is usually much smaller than
that of $\theta<6^\circ$.
The incident high-energy photons ($>5$ MeV) can deposit a large amount of energy in the low-energy band as the result of Compton scattering.
The photons incident with $\theta>6^\circ$ will suffer serious scattering and absorption (Panel (b) and Panel (e) in Figure \ref{RSPjpg})
that can affect the energy resolution of the detectors.
Figure \ref{ARF_allsatjpg} shows the effective areas of \emph{HXMT}/CsI and other gamma-ray instruments operating in orbit currently \citep{Stamatikos2008,stamatikos2009crosscalibration,2013ICRC...33.2948Y,astrosathandbook}.

The simulative spectral analyses are performed to investigate
the independent spectral capabilities of \emph{HXMT}/CsI in different incident directions.
The spectral analyses are performed with the Band GRB model \citep{Band1993BATSE}:
\begin{equation}\label{model:Band}
{f\rm{_{Band}}}(E) = A
\left\{
    \begin{array}{ll}
         {\left( {\frac{E}{{100\;\rm{keV}}}} \right)^\alpha }\rm{exp} \left[ { - \frac{{\left( {\alpha  + 2)\;\emph{E}} \right.}}{{{\emph{E}_{peak}}}}} \right], E < \frac{{\left( {\alpha  - \beta } \right)\;{E_{\rm{peak}}}}}{{\alpha  + 2}}, \hfill \\
         {\left( {\frac{E}{{100\;\rm{keV}}}} \right)^\beta }\rm{exp} \left( {\beta {\text{ - }}\alpha } \right){\left[ {\frac{{\left( {\alpha  - \beta } \right)\;{\emph{E}_{peak}}}}{{100\;\rm{keV}\;\left( {\alpha  + 2} \right)}}} \right]^{\alpha  - \beta }}, E\geqslant \frac{{\left( {\alpha  - \beta } \right)\;{E_{\rm{peak}}}}}{{\alpha  + 2}}, \hfill \\
    \end{array}
\right.,
\end{equation}
where $A$ is the amplitude ($\rm cts~s^{-1}~cm^{-2}~keV^{-1}$), $\alpha$ and $\beta$ are the low- and high-energy spectral indices, respectively, and $E\rm_{peak}$ is the $\nu F_{\nu}$ peak energy.
Incident directions $(\theta,\phi) = (0^\circ,0^\circ)$, $(60^\circ,0^\circ)$, $(60^\circ,90^\circ)$, $(150^\circ,0^\circ)$, and $(150^\circ,90^\circ)$ are selected to perform
the simulative spectral fittings. For each $(\theta,\phi)$, we simulate GRB with three spectral models (Table~\ref{three_type_Band}), each characterized as well with different average fluxes and durations (see Table \ref{table1simuPara}).
We take the background spectral shapes of typical observation events to generate the background data for simulation, i.e., GRB 170626A for NG mode and GRB 181028A for LG mode.

\begin{figure*}[htbp]
    \centering
    \subfigure[NG, all parameters free]{
    \begin{minipage}[t]{0.5\linewidth}
    \centering
    \includegraphics[width=1\linewidth]{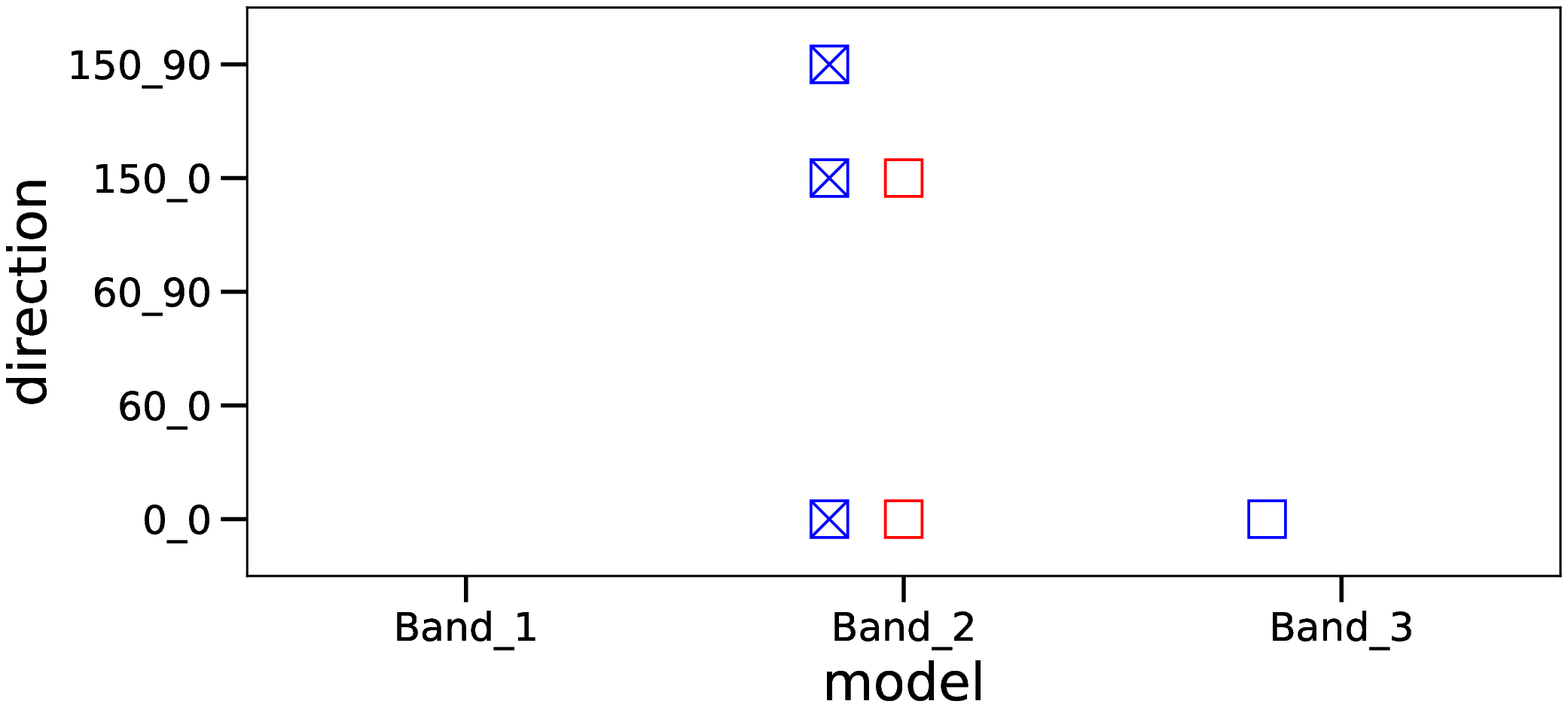}
    \end{minipage}\label{simu_untie_NG}%
    }%
    \subfigure[LG, all parameters free]{
    \begin{minipage}[t]{0.5\linewidth}
    \centering
    \includegraphics[width=1\linewidth]{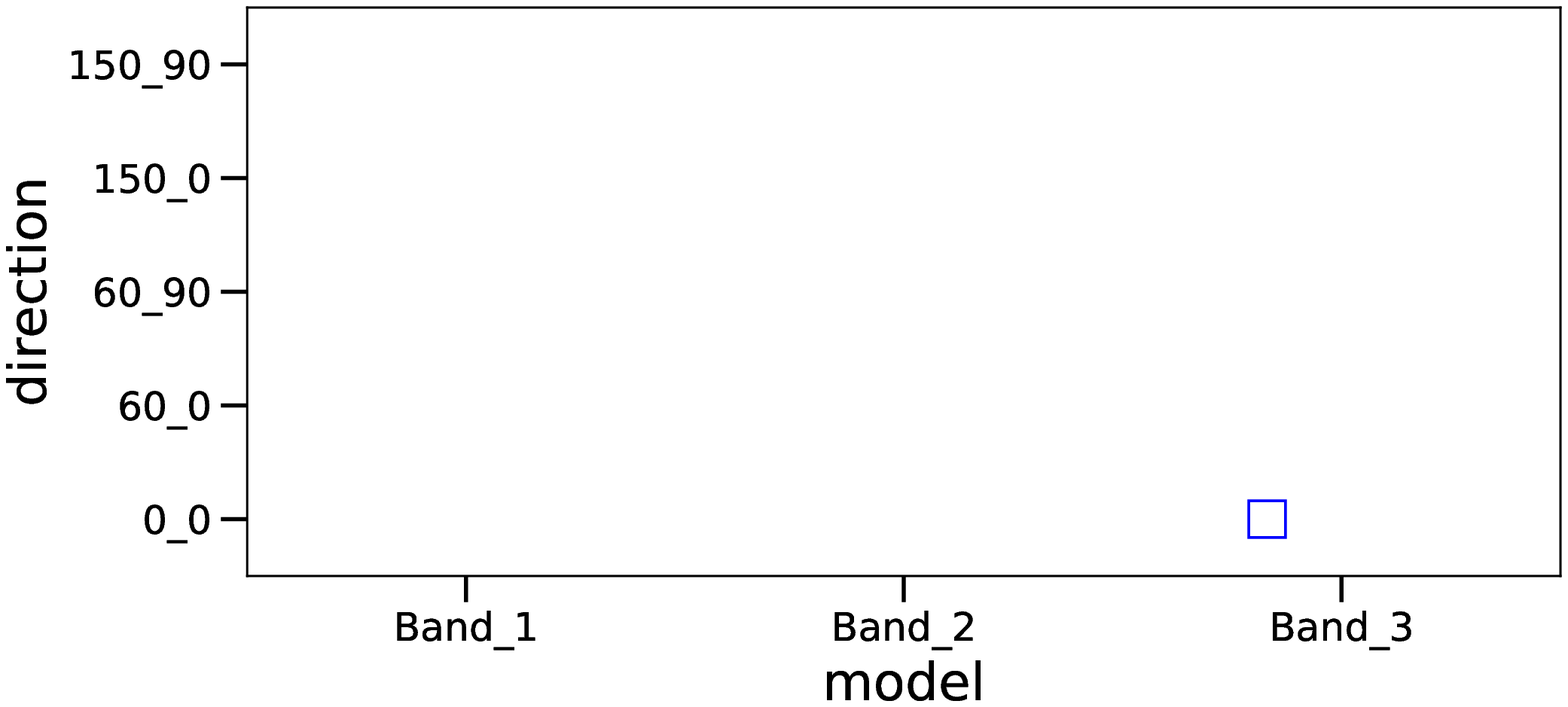}
    \end{minipage}\label{simu_untie_LG}%
    }%
    \quad
    \subfigure[NG, $\alpha$ frozen]{
    \begin{minipage}[t]{0.5\linewidth}
    \centering
    \includegraphics[width=1\linewidth]{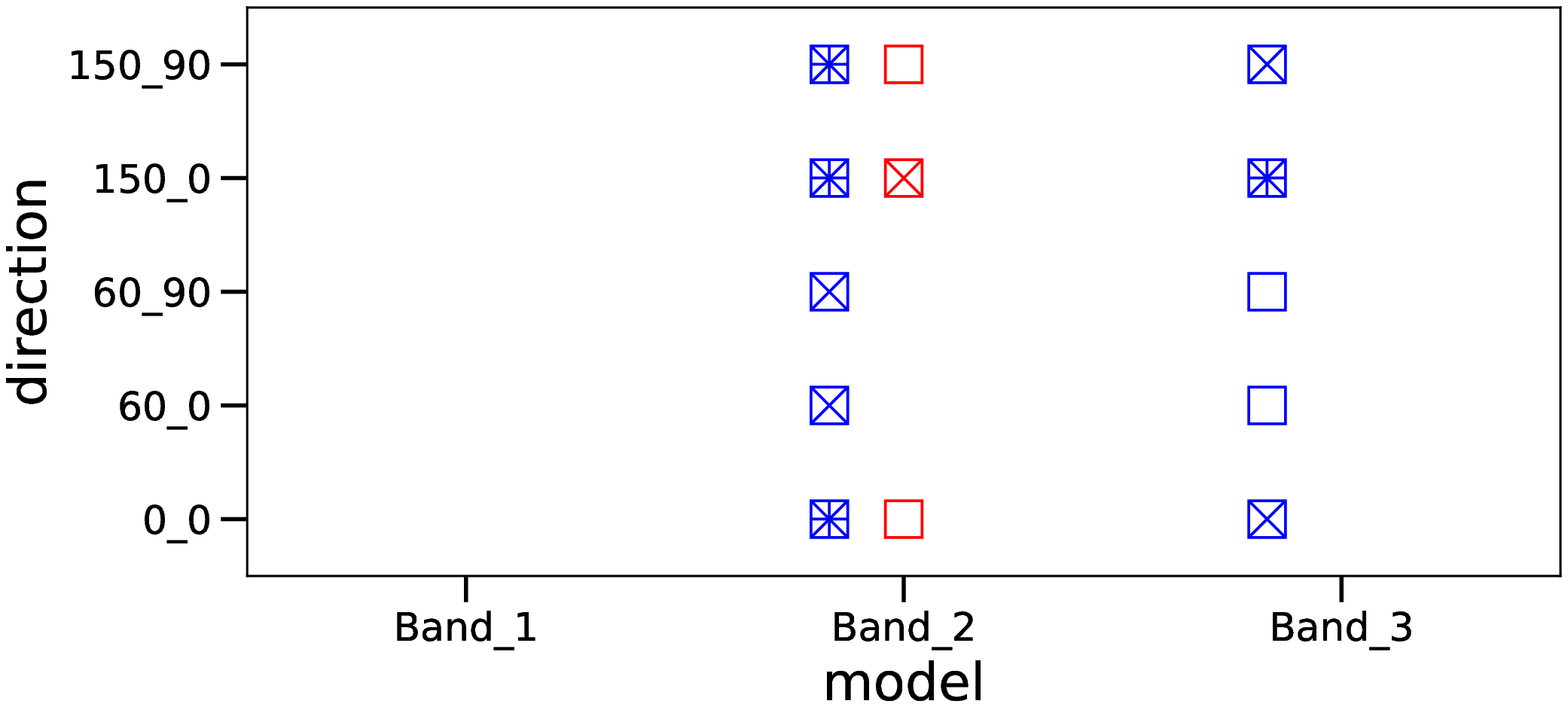}
    \end{minipage}\label{simu_alpha_NG}%
    }%
    \subfigure[LG, $\alpha$ frozen]{
    \begin{minipage}[t]{0.5\linewidth}
    \centering
    \includegraphics[width=1\linewidth]{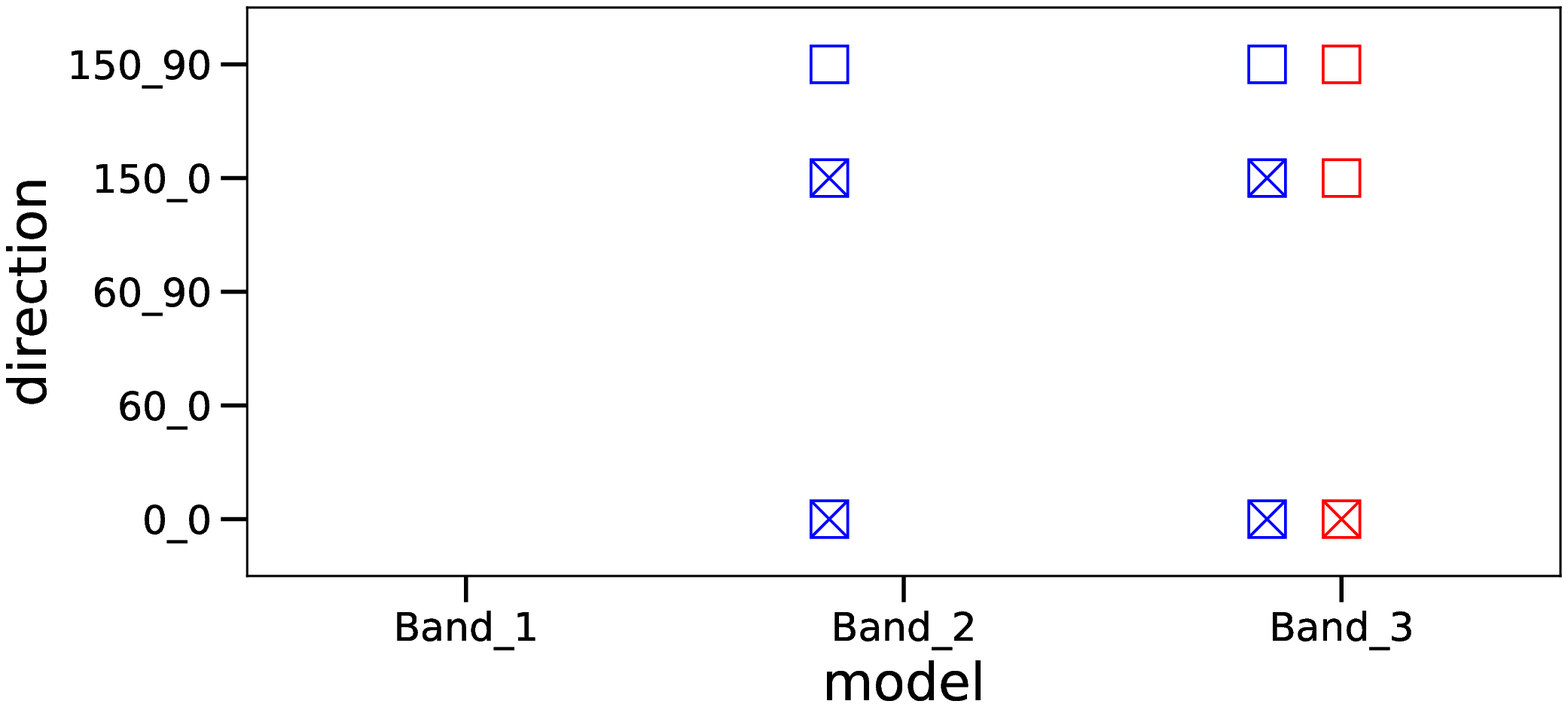}
    \end{minipage}\label{simu_alpha_LG}%
    }%
    \quad
    \subfigure[NG, $\beta$ frozen]{
    \begin{minipage}[t]{0.5\linewidth}
    \centering
    \includegraphics[width=1\linewidth]{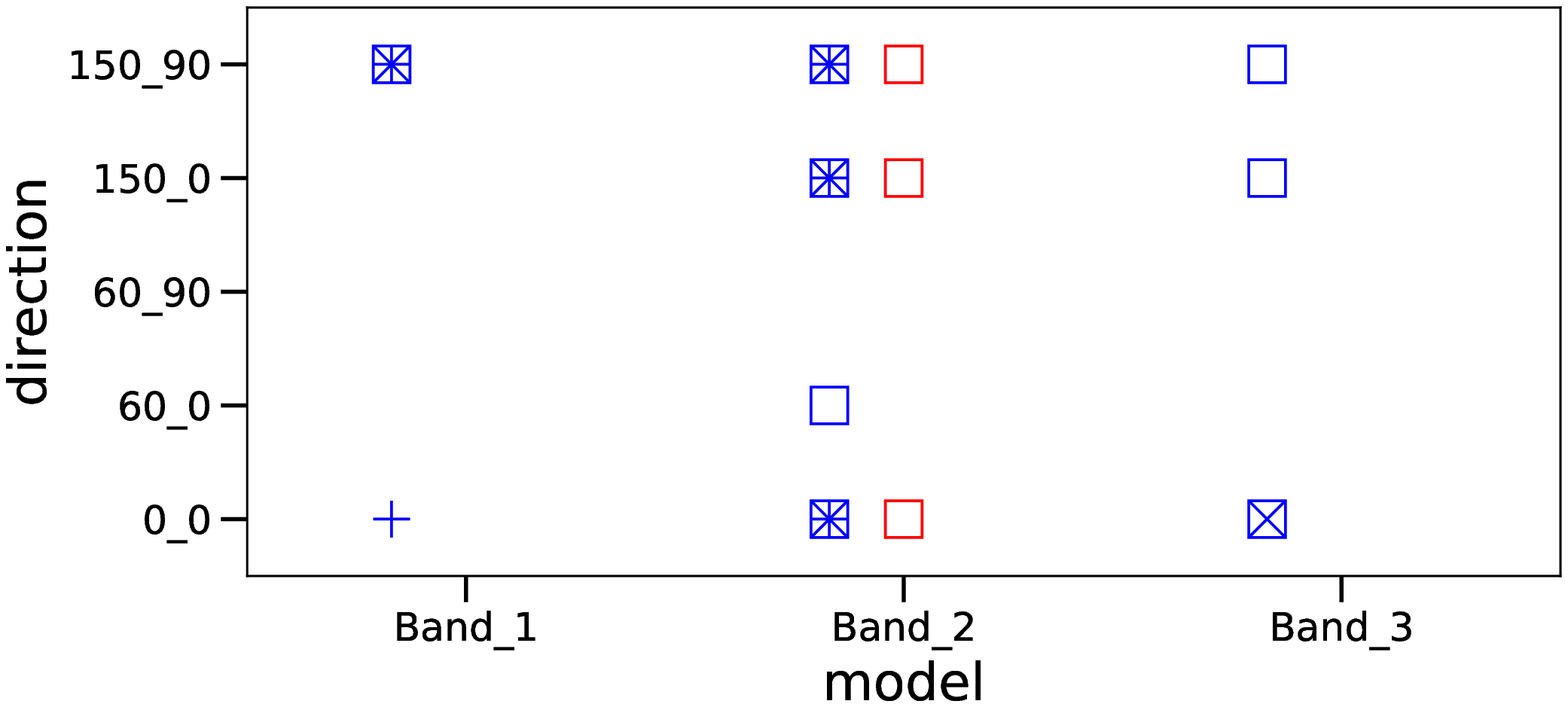}
    \end{minipage}\label{simu_beta_NG}%
    }%
    \subfigure[LG, $\beta$ frozen]{
    \begin{minipage}[t]{0.5\linewidth}
    \centering
    \includegraphics[width=1\linewidth]{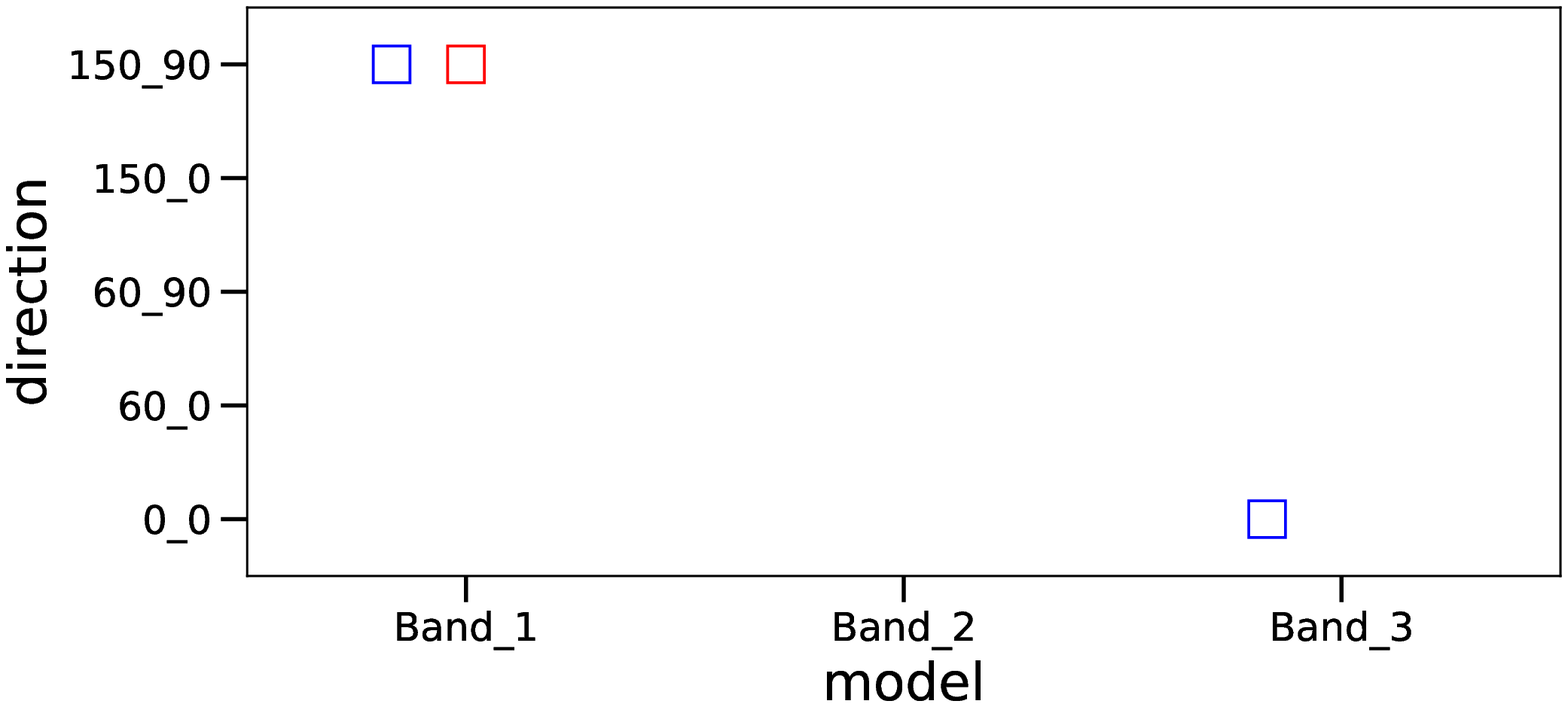}
    \end{minipage}\label{simu_beta_LG}%
    }%
    \centering
    \caption{Results of the simulative spectral fitting with the Band GRB model in NG and LG mode.
	The spectral fitting with all parameters free, $\alpha$ frozen and $\beta$ frozen are shown from the top to the bottom.
	The points in these panels indicate that all the fitting parameters can be well constrained.
	The X-axis is the fitting model (Table~\ref{three_type_Band}) and the Y-axis is the setting incident direction ($\theta,\phi$).
	The blue and red colors refer that the flux are $10^{-5}$ and $10^{-6}$ erg$\;$cm$^{-2}$$\;$s$^{-1}$, respectively. All the parameters of the simulated spectra with flux = $10^{-7}$ erg$\;$cm$^{-2}$$\;$s$^{-1}$
can not be constrained.
	The burst durations $t=1~{\rm s},10~{\rm s},100~{\rm s}$ are shown as `+', `$\times$', `$\square$', respectively.}\label{simuresultsjpg}
\end{figure*}

For each fitting, all the 18 simulated \emph{HXMT}/CsI spectra and responses are merged first and then the merged spectrum is fitted with the merged response.
The detailed results of the simulative spectral analyses are shown in Figure~\ref{simuresultsjpg}.
It is obvious that the spectral capability of \emph{HXMT}/CsI in the high-energy band is much better than that in the low energy range for both the NG and LG modes.
Due to limitation of the low efficiency in low energy band, \emph{HXMT}/CsI can not constrain all the spectral parameters in the whole energy band for most of the GRBs.
For the NG mode, the spectral parameters in both the low- and high-energy bands can be well constrained only for the GRBs with both harder spectra and larger fluxes (Figure \ref{simu_untie_NG}).
For the LG mode, the spectral parameters can not be constrained for all the simulated spectra once setting free all the parameters (Figure \ref{simu_untie_LG}).
Therefore, we perform the spectral fitting under a fixed low-energy spectral index $\alpha$.
The result shows that \emph{HXMT}/CsI can well constrain the spectrum for GRB with a harder spectral shape, a relatively higher flux and a longer duration.
Thus the GRB research of \emph{HXMT}/CsI shall be carried out jointly with other missions which provide supplementary measurements at lower energy bands.

\begin{table}[!htbp]
\caption{Three types of Band GRB model in simulated spectral analysis.}
\begin{center}
\renewcommand\arraystretch{1.0}
\begin{tabular}{p{1.2cm}<{\centering}p{1.2cm}<{\centering}p{1.2cm}<{\centering}p{2.4cm}<{\centering}}
\hline
model& $\alpha$& $\beta$& $E\rm_{peak}$ (keV)\\
\hline
Band\_1& -1.9& -3.7& 70\\
Band\_2& -1.0& -2.3& 230\\
Band\_3& 0.0& -1.5& 1000\\
\hline
\end{tabular}
\end{center} \label{three_type_Band}
\end{table}

\begin{table*}[!htbp]
\caption{Parameters in the simulative spectral analysis of \emph{HXMT}/CsI.}
\begin{center}
\renewcommand\arraystretch{1.0}
\begin{tabular}{ccccccc}
\hline
Incident Direction & ($0^\circ$, $0^\circ$)& ($60^\circ $, $0^\circ$)& ($60^\circ $, $90^\circ$)& ($150^\circ$, $0^\circ$)& ($150^\circ$, $90^\circ$)& \\
Spectral Type$^{*}$& \multicolumn{2}{l}{Band\_1}& \multicolumn{2}{l}{Band\_2}& \multicolumn{2}{l}{Band\_3}\\
Flux (erg$\;$cm$^{-2}$$\;$s$^{-1}$)$^{\#}$& \multicolumn{2}{l}{$10^{-5}$ } & \multicolumn{2}{l}{$10^{-6}$} & \multicolumn{2}{l}{$10^{-7}$} \\
Burst Duration (s)& \multicolumn{2}{l}{1}& \multicolumn{2}{l}{10}& \multicolumn{2}{l}{100}\\
\hline
\end{tabular}
\end{center} \label{table1simuPara}
\footnotesize{$^{*}$ The parameters of all the spectral type are shown in Table \ref{three_type_Band}.} \\
\footnotesize{$^{\#}$ The energy range is $20-40,000$~keV.}
\end{table*}

\section{Joint Spectral Analyses with $Fermi$/GBM, $Swift$/BAT, Konus-\emph{Wind}}
\label{sec:Joint-Fitting}

\subsection{$Fermi$/GBM, $Swift$/BAT and Konus-\emph{Wind}}
The instrumental response of \emph{HXMT}/CsI can be cross calibrated with GRB observations carried out simultaneously by other in-orbit telescopes (\emph{Fermi}/GBM, \emph{Swift}/BAT, Konus-\emph{Wind}).

The \emph{Fermi} Gamma-Ray Space Telescope is composed of two science instruments, i.e., the Large Area Telescope (LAT) and the Gamma-Ray Burst Monitor (GBM).
The GBM comprises 12 thallium activated sodium iodide (NaI(Tl)) scintillation detectors and two bismuth germanate (BGO) scintillation detectors. The NaI(Tl)
detectors work at 8~keV $-$ 1~MeV and the BGO detectors cover an energy range of $\sim200$~keV to $\sim4$~MeV. The NaI(Tl) crystal is shaped with a
diameter of $12.7$ cm and a thickness of $1.27$ cm, whereas the BGO crystal has a diameter of $12.7$ cm \citep{Meegan2009}.
For each burst, GBM provides three types of science data: CTIME, CSPEC and Time-Tagged Events (TTE) data \citep{Paciesas2012}. We use CSPEC or TTE data for spectral analysis.
In the joint spectral fitting, only two NaI detectors and one BGO detector with large count rate are considered.

The Burst Alert Telescope (BAT) is one of the three instruments onboard the \emph{Swift} MIDEX spacecraft. BAT is a coded-aperture instrument with a wide field of view. The BAT detector plane is composed of 32,768 pieces of CdZnTe, and the coded-aperture mask owns $\sim52000$ plumbum tiles separated by one meter from the detector plane. BAT works at $10-150$~keV and has an energy resolution as $\sim7$~keV \citep{Barthelmy2003}.

The Konus-\emph{Wind} (KW) is a gamma-ray spectrometer for temporal and spectral analyses of GRB.
It consists of two identical omnidirectional NaI(Tl) detectors (S1 and S2), each with an effective area of about
$80-160$~cm$^2$ for photons of different incident energies and directions \citep{Aptekar1995Konus,Sakamoto2011}.
KW joint observations upon GRB 181028A are considered in our cross calibration analysis, in which the two KW spectra are generated by the KW team in energy bands of
$30-1500$~keV (PHA1) and $0.5-18$~MeV (PHA2), respectively.

\subsection{Joint Spectral Analyses}
Ten bright GRBs observed simultaneously with \emph{HXMT}/CsI and other instruments (\emph{Fermi}/GBM, \emph{Swift}/BAT, Konus-\emph{Wind})
are selected to perform the joint spectral analyses.
The time-averaged energy spectrum is generated for each telescope, with light
aberration between different instruments properly
fitted with the Band GRB model.
In the joint fitting, an adjustable parameter $A$ is introduced for each specific instrument to account partially for the uncertainty in effective area calibration.
The detailed results of the joint fittings are shown in Table~\ref{table2}.

\begin{figure*}[t]
    \centering
    \includegraphics[width=1\textwidth]{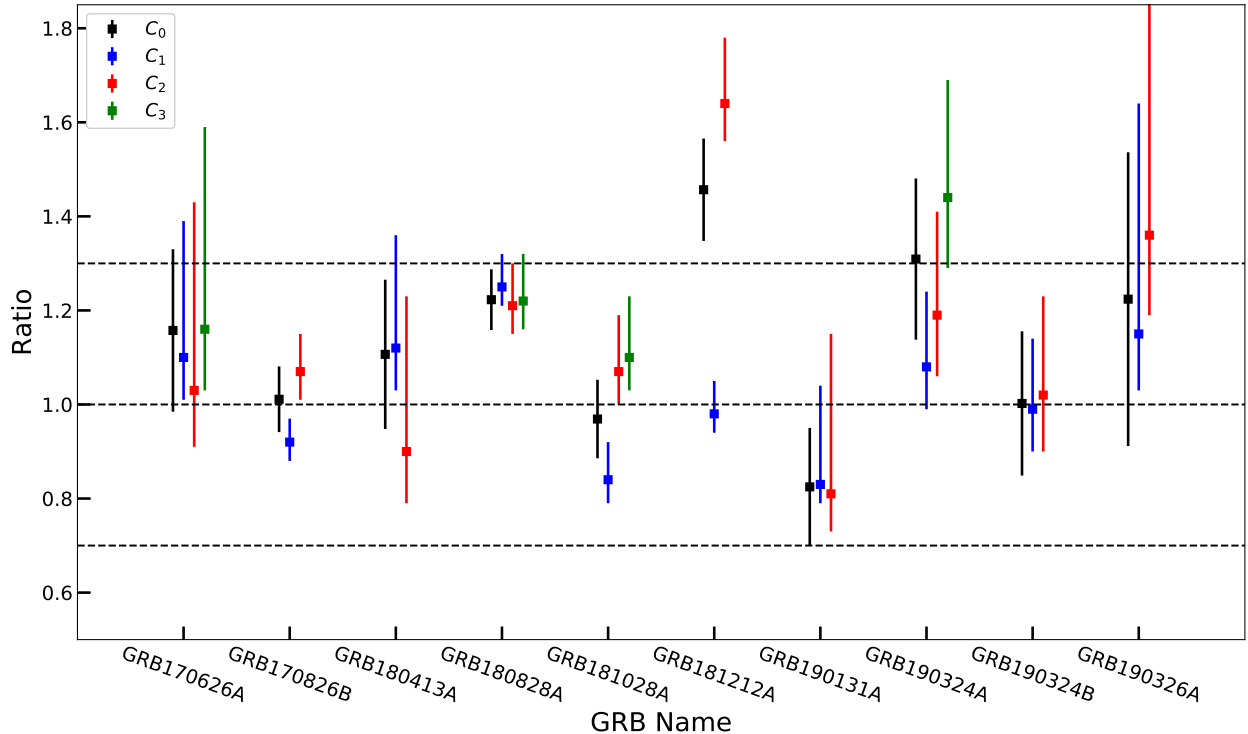}\\
  \caption{Ratios of the amplitudes of \emph{HXMT}/CsI to these of other instruments in the joint spectral fittings. The black, blue, red, and green points are $C_0$, $C_1$, $C_2$, and $C_3$, respectively.}\label{constantjpg}
\end{figure*}

The results show that the amplitudes obtained by all the instruments are consistent
for all GRBs except GRB~181212A.
For GRB~181212A, the \emph{HXMT}/CsI amplitude is similar to \emph{Fermi}/GBM BGO detector,
but deviates significantly from \emph{Fermi}/GBM NaI detector.
It is worth noting that such an amplitude deviation still holds even if only the \emph{Fermi}/GBM data are considered.
This may suggest the possible shortages either in the adopted model or the understanding of the
Fermi/GBM response.
We take the following parameters to describe the consistency of \emph{HXMT}/CsI with other instruments,
\begin{equation}\label{func:C0}
\begin{split}
 {C_{0} = \frac{A_{\rm{H}}}{A_{\rm{m}}}},~{C_{1} = \frac{A_{\rm{H}}}{A_{\rm{FB}}}},~{C_{2} = \frac{A_{\rm{H}}}{A_{\rm{FN}}}},~\\
{C_{3}} =
\left\{
    \begin{array}{ll}
     \frac{A_{\rm{H}}}{A_{\rm{S}}} , \rm{\;for\;GRBs\;except\;GRB~181028A}, \hfill \\
     \hfill \\
     \frac{A_{\rm{H}}}{A_{\rm{K}}} , \rm{\;for\;GRB~181028A}, \hfill \\
    \end{array}
\right.,
\end{split}
\end{equation}
where $A_{\rm{H}}$, $A_{\rm{FB}}$, $A_{\rm{FN}}$, $A_{\rm{S}}$ and $A_{\rm{K}}$ refer the amplitudes of \emph{HXMT}/CsI,
\emph{Fermi}/GBM BGO, \emph{Fermi}/GBM NaI, \emph{Swift}/BAT and Konus-\emph{Wind}, respectively; $A_{\rm{m}}$ refers the
weighted average of $A_{\rm{FB}}$, $A_{\rm{FN}}$, $A_{\rm{S}}$ and $A_{\rm{K}}$.
As shown in Figure~\ref{constantjpg}, $C_0$ is slightly greater than unity.
A maximum likelihood approach \citep{Liao_2013} is applied to
$C_0$ for investigating the difference between \emph{HXMT}/CsI and other instruments. In order to avoid the potential uncertainty described above, GRB~181212A is excluded from the sample.
The result shows that $A_{\rm{H}}$ is systematically higher by $7.0\pm8.8\%$ than the amplitudes of other instruments,
which is not significant due to the small number of the sample.

\begin{figure*}[t]
    \centering
    \subfigure{
    \begin{minipage}[t]{0.5\linewidth}
    \centering
    \includegraphics[width=1\linewidth]{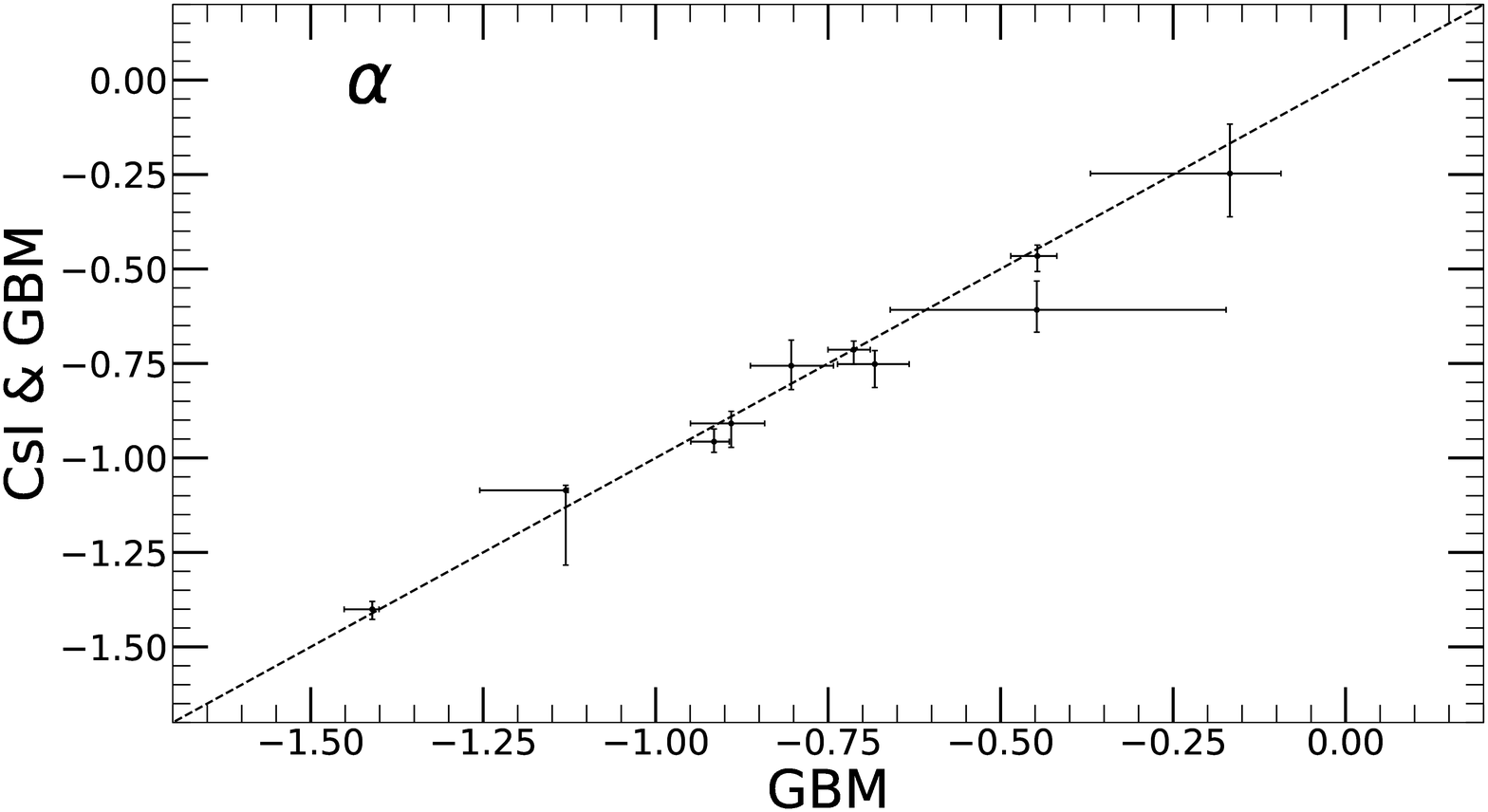}
    \end{minipage}%
    }%
    \subfigure{
    \begin{minipage}[t]{0.5\linewidth}
    \centering
    \includegraphics[width=1\linewidth]{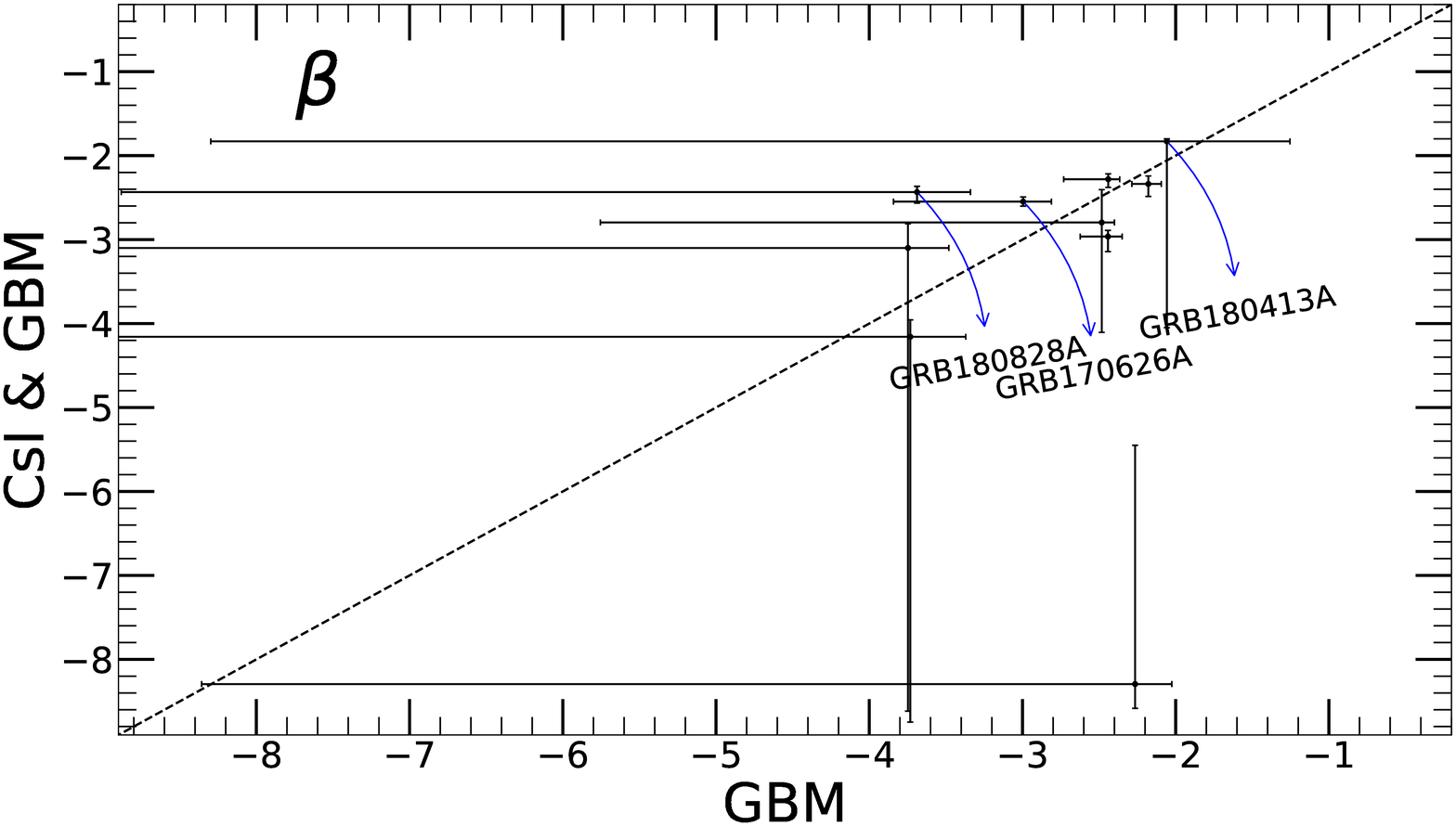}
    \end{minipage}%
    }%

    \quad
    \subfigure{
    \begin{minipage}[t]{0.5\linewidth}
    \centering
    \includegraphics[width=1\linewidth]{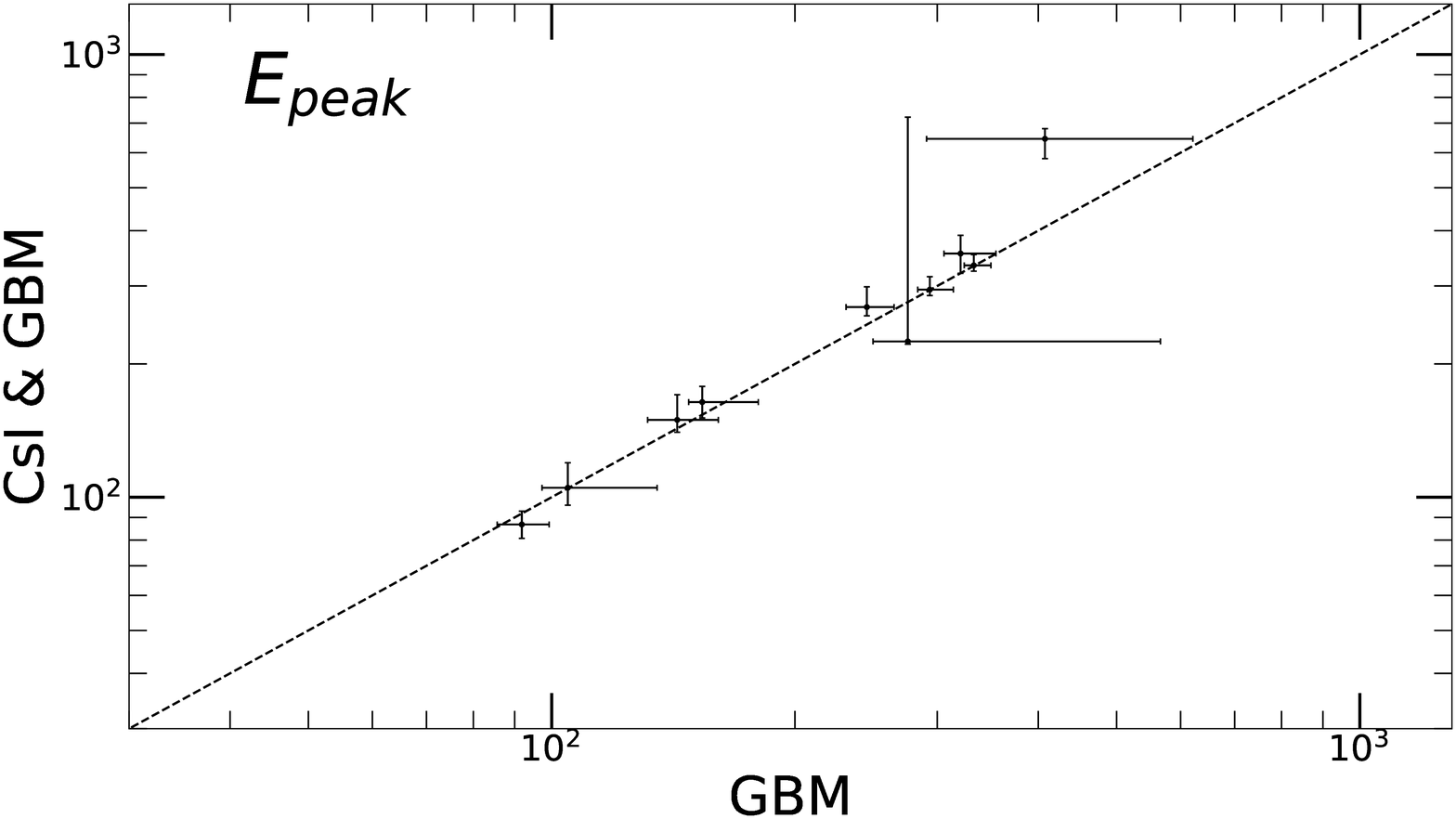}
    \end{minipage}%
    }%
    \subfigure{
    \begin{minipage}[t]{0.5\linewidth}
    \centering
    \includegraphics[width=1\linewidth]{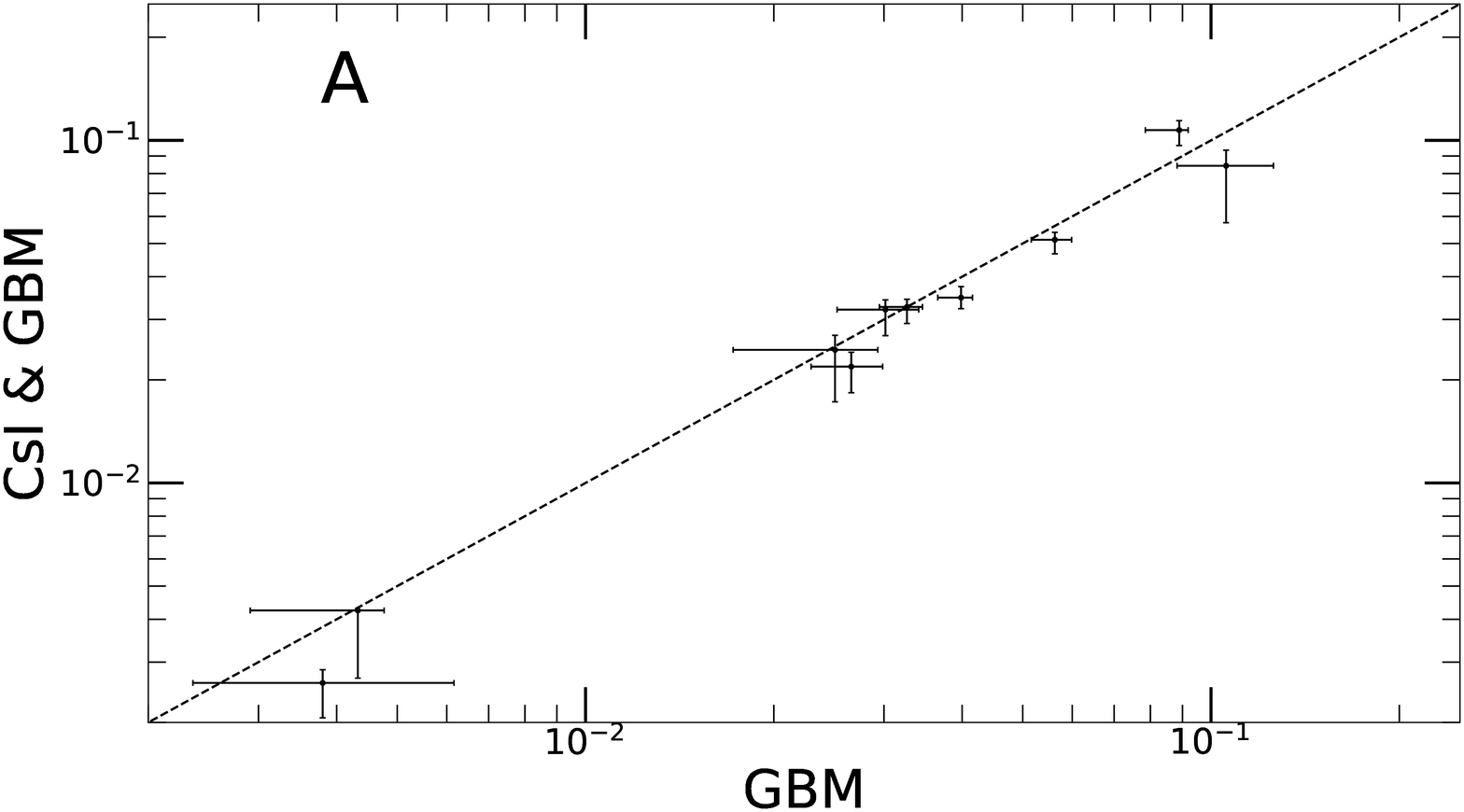}
    \end{minipage}%
    }%
    \centering
    \caption{Comparisons of $Fermi$/GBM independent spectral fitting and joint \emph{HXMT}/CsI-\emph{Fermi}/GBM spectral fitting. As shown in the top-right panel, the error bars in X-axis are smaller than these in Y-axis, especially for GRB 170626A, GRB 180413A and GRB 180828A.}\label{parameter}
\end{figure*}

\cite{Guidorzi2019} once demonstrates that \emph{HXMT}/CsI can be very helpful, thanks to its large effective area at soft gamma-rays (e.g., $200-2000$~keV in LG mode, Figure~\ref{ARFjpg}), in constraining the GRB spectrum via joint analysis with other instruments.
They find that the the high-energy spectral index beta can be largely improved by adding additionally \emph{HXMT}/CsI data into \emph{Fermi}/GBM.
For some GRBs like GRB~180413A \& GRB~180828A, the high-energy spectral indices $\beta$ can only be measured to a precision with systematic error $\sim0.1$ by \emph{Fermi}/GBM once \emph{HXMT}/CsI data are included.
The other parameters $\alpha$, $E\rm_{peak}$ and $A$ resulted from the joint \emph{HXMT}/CsI-\emph{Fermi}/GBM fittings are consistent with those given by taking \emph{Fermi}/GBM alone (Figure \ref{parameter}), suggesting that the advantage of \emph{HXMT}/CsI is mainly in the soft gamma-ray band.

\begin{table*}[!htbp]
\caption{Results of the joint spectral fittings.}
\begin{center}
\setlength{\tabcolsep}{1mm}
\resizebox{16.2cm}{2.5cm}{
\renewcommand\arraystretch{1.2}
\begin{tabular}{ccccccccccc}
\hline
GRB Name & Direction ($\theta,\phi$) & $\alpha$ & $\beta$ & $E_{\rm peak}~({\rm keV})$ & $A_{\rm H}$ & $A_{\rm FB}$ & $A_{\rm FN}$ $^*$ & $A_{\rm{S}}$ & $A_{\rm K}$ & $\chi^{2}$/d.o.f.\\
\hline

170626A& (113$^\circ$, 212$^\circ$)& $-0.79$$^{+0.11}_{-0.03}$ &$-2.51$$^{+0.07}_{-0.04}$ &$82.3$$^{+3.1}_{-8.6}$ &$9.39$$^{+2.48}_{-0.77}$ &$8.54$$^{+1.86}_{-1.81}$ & $9.12$$^{+1.06}_{-3.54}$ &$8.10$$^{+0.16}_{-0.29}$ &-- &662/674\\

170826B& (53$^\circ$, 72$^\circ$)& $-0.96$$^{+0.03}_{-0.03}$ &$-2.28$$^{+0.06}_{-0.10}$ &$355.1$$^{+35.4}_{-34.7}$ &$3.20$$^{+0.17}_{-0.14}$ &$3.48$$^{+0.27}_{-0.25}$ & $2.99$$^{+0.17}_{-0.22}$ &-- &-- &636/600\\

180413A& (112$^\circ$, 108$^\circ$)& $-1.09$$^{+0.01}_{-0.20}$ &$-1.83$$^{+0.03}_{-2.22}$ &$224.8$$^{+497.1}_{-3.1}$ &$0.47$$^{+0.10}_{-0.04}$ &$0.42$$^{+0.01}_{-0.16}$ & $0.52$$^{+0.06}_{-0.19}$ &-- &-- &564/586\\

180828A& (39$^\circ$, 355$^\circ$)& $-0.50$$^{+0.03}_{-0.04}$ &$-2.46$$^{+0.08}_{-0.14}$ &$345.1$$^{+19.9}_{-15.0}$ &$6.15$$^{+0.34}_{-0.20}$ &$4.92$$^{+0.32}_{-0.42}$ & $5.08$$^{+0.25}_{-0.38}$ &$5.04$$^{+0.16}_{-0.29}$ &-- &655/643\\

181028A& (130$^\circ$, 107$^\circ$)& $-0.71$$^{+0.02}_{-0.03}$ &$-4.38$$^{+0.47}_{-4.46}$ &$296.0$$^{+11.7}_{-7.0}$ &$2.72$$^{+0.26}_{-0.16}$ &$3.24$$^{+0.14}_{-0.24}$ & $2.54$$^{+0.17}_{-0.29}$ &-- &$2.47$$^{+0.16}_{-0.29}$ &661/717\\

181212A& (142$^\circ$, 77$^\circ$)& $-1.40$$^{+0.02}_{-0.03}$ &$-2.96$$^{+0.07}_{-0.18}$ &$105.0$$^{+14.6}_{-9.1}$ &$10.50$$^{+0.75}_{-0.43}$ &$10.71$$^{+0.70}_{-1.06}$ & $6.40$$^{+0.31}_{-0.55}$ &-- &-- &661/586\\

190131A& (151$^\circ$, 175$^\circ$)& $-0.61$$^{+0.08}_{-0.06}$ &$-8.29$$^{+2.84}_{-0.29}$ &$644.9$$^{+34.9}_{-63.0}$ &$0.22$$^{+0.05}_{-0.01}$ &$0.26$$^{+0.02}_{-0.05}$ & $0.27$$^{+0.03}_{-0.11}$ &-- &-- &472/586\\

190324A& (68$^\circ$, 227$^\circ$)& $-0.96$$^{+0.05}_{-0.06}$ &$-2.29$$^{+0.10}_{-0.15}$ &$149.5$$^{+21.2}_{-16.0}$ &$3.22$$^{+0.48}_{-0.27}$ &$2.98$$^{+0.41}_{-0.40}$ & $2.70$$^{+0.30}_{-0.50}$ &$2.24$$^{+0.16}_{-0.29}$ &-- &664/660\\

190324B& (156$^\circ$, 188$^\circ$)& $-0.75$$^{+0.04}_{-0.06}$ &$-3.10$$^{+0.29}_{-5.52}$ &$268.8$$^{+29.9}_{-11.9}$ &$2.17$$^{+0.33}_{-0.20}$ &$2.19$$^{+0.22}_{-0.35}$ & $2.13$$^{+0.25}_{-0.44}$ &-- &-- &614/589\\

190326A& (123$^\circ$, 22$^\circ$)& $-0.25$$^{+0.13}_{-0.11}$ &$-2.80$$^{+0.39}_{-1.31}$ &$164.0$$^{+13.9}_{-13.0}$ &$2.82$$^{+1.20}_{-0.29}$ &$2.45$$^{+0.25}_{-0.72}$ & $2.07$$^{+0.26}_{-1.08}$ &-- &-- &552/589\\

\hline
\end{tabular}
}
\end{center} \label{table2}
\footnotesize{Note. The units of $A_{\rm H}$, $A_{\rm FB}$, $A_{\rm FN}$, $A_{\rm{S}}$ and $A_{\rm K}$ are $\rm 10^{-2}~cts~s^{-1}~keV^{-1}~cm^{-2}$.\\}
\footnotesize{$^*$ Both the two \emph{Fermi}/GBM NaI detectors used in the joint fitting have the same fitting parameters.}
\end{table*}

\begin{figure*}[htbp]
    \centering
    \subfigure[GRB~170626A, NG mode]{
    \begin{minipage}[t]{0.5\linewidth}
    \centering
    \includegraphics[width=2.5in]{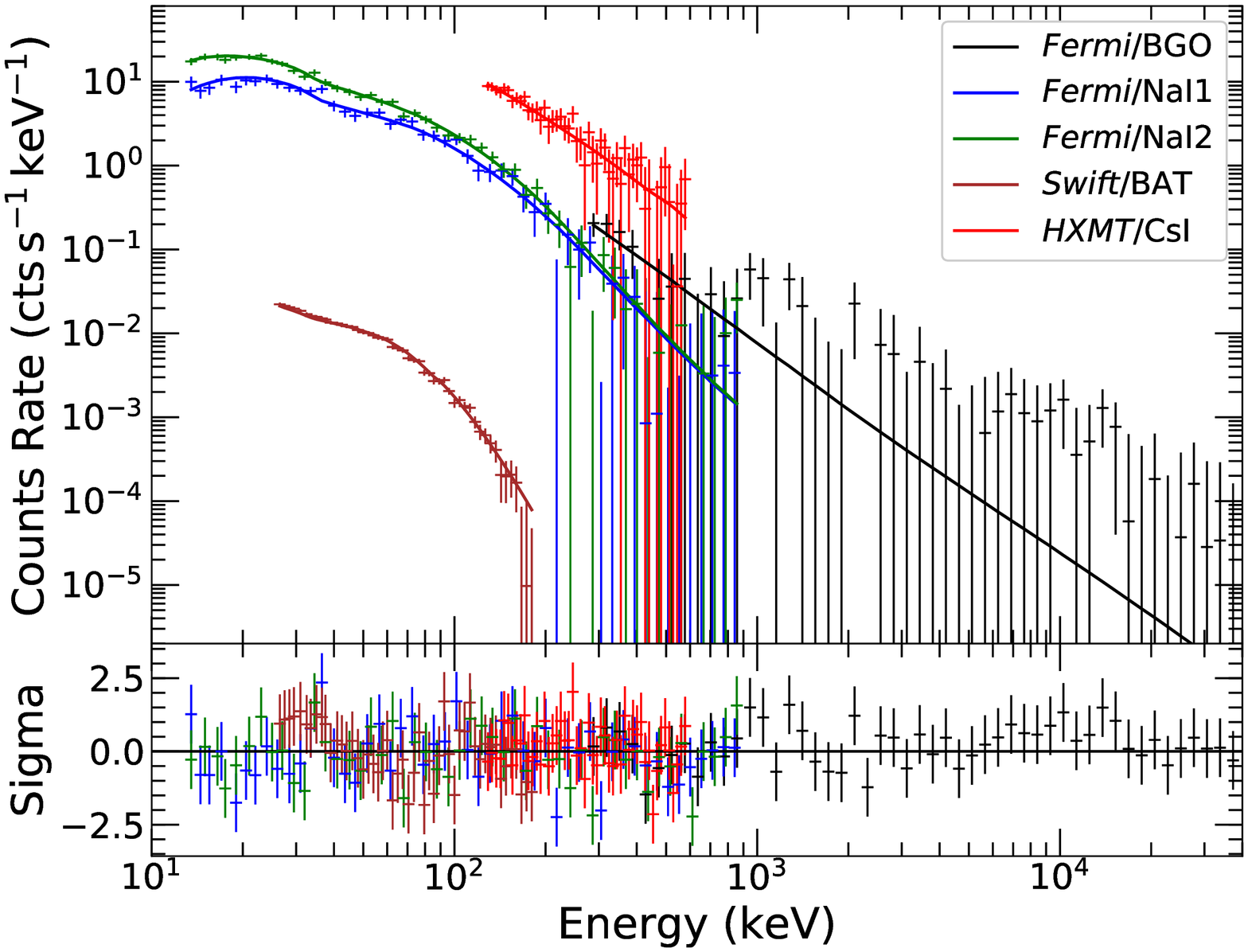}
    \end{minipage}%
    }%
    \subfigure[GRB~170826B, NG mode]{
    \begin{minipage}[t]{0.5\linewidth}
    \centering
    \includegraphics[width=2.5in]{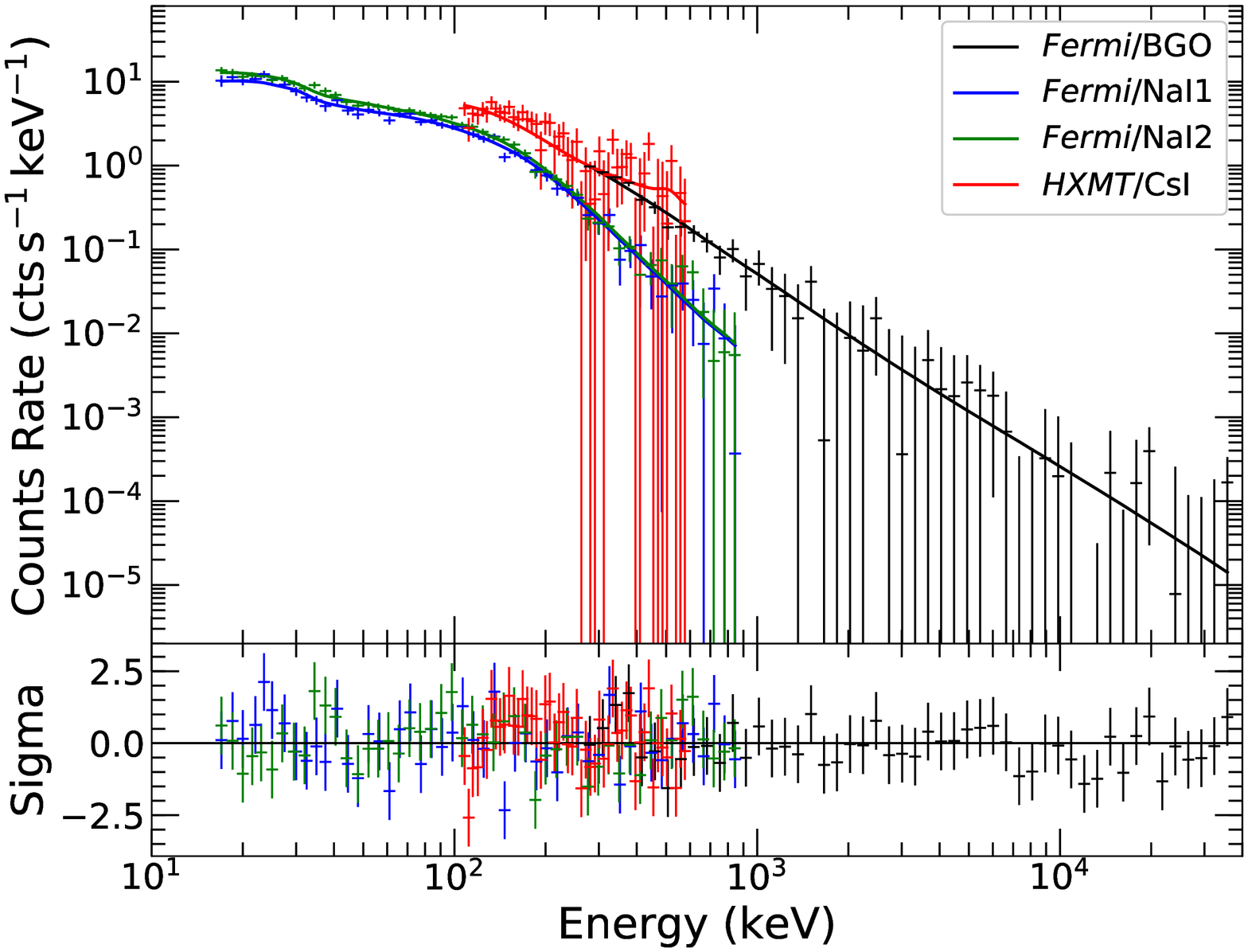}
    \end{minipage}%
    }%

    \quad
    \subfigure[GRB~180413A, NG mode]{
    \begin{minipage}[t]{0.5\linewidth}
    \centering
    \includegraphics[width=2.5in]{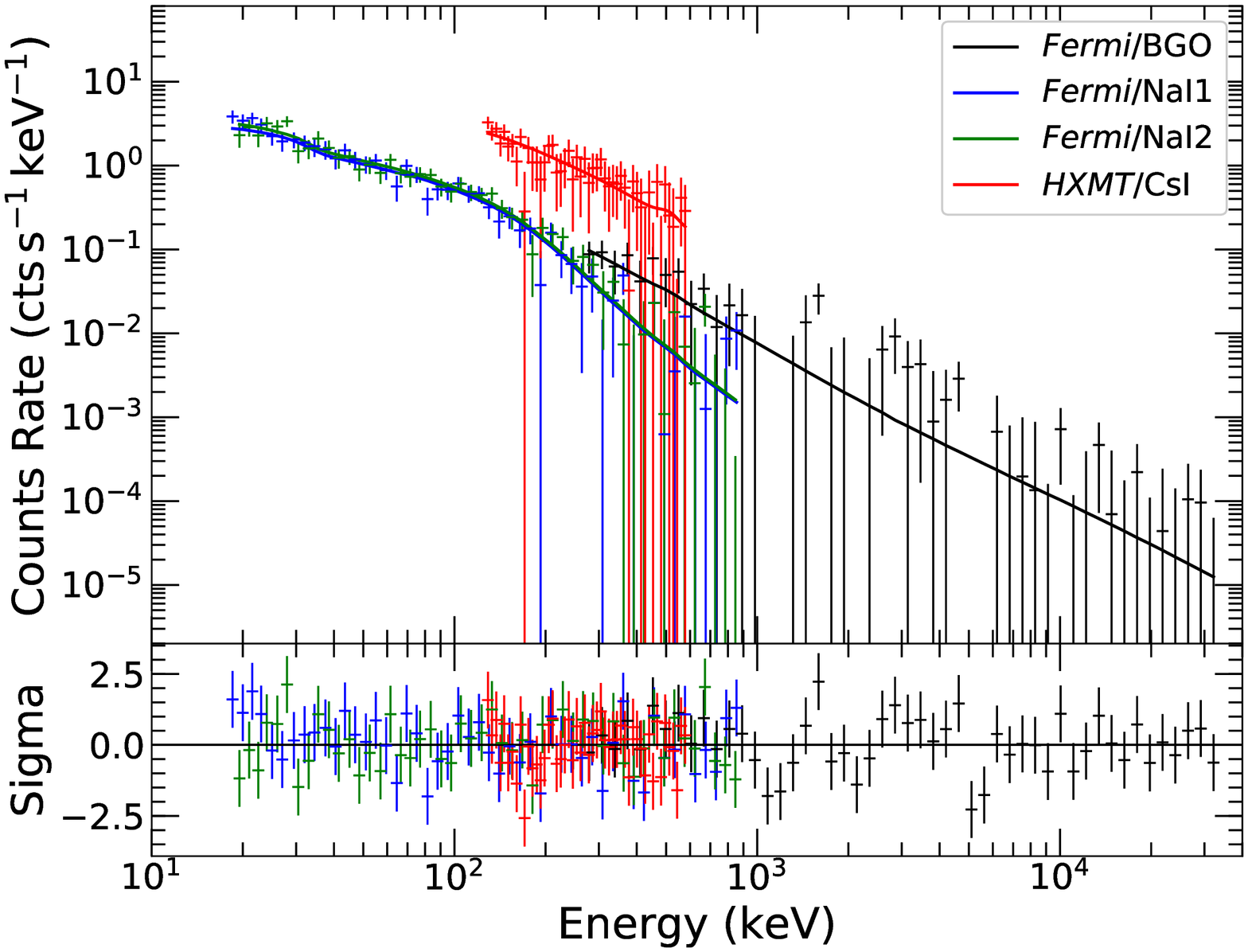}
    \end{minipage}%
    }%
    \subfigure[GRB~180828A, NG mode]{
    \begin{minipage}[t]{0.5\linewidth}
    \centering
    \includegraphics[width=2.5in]{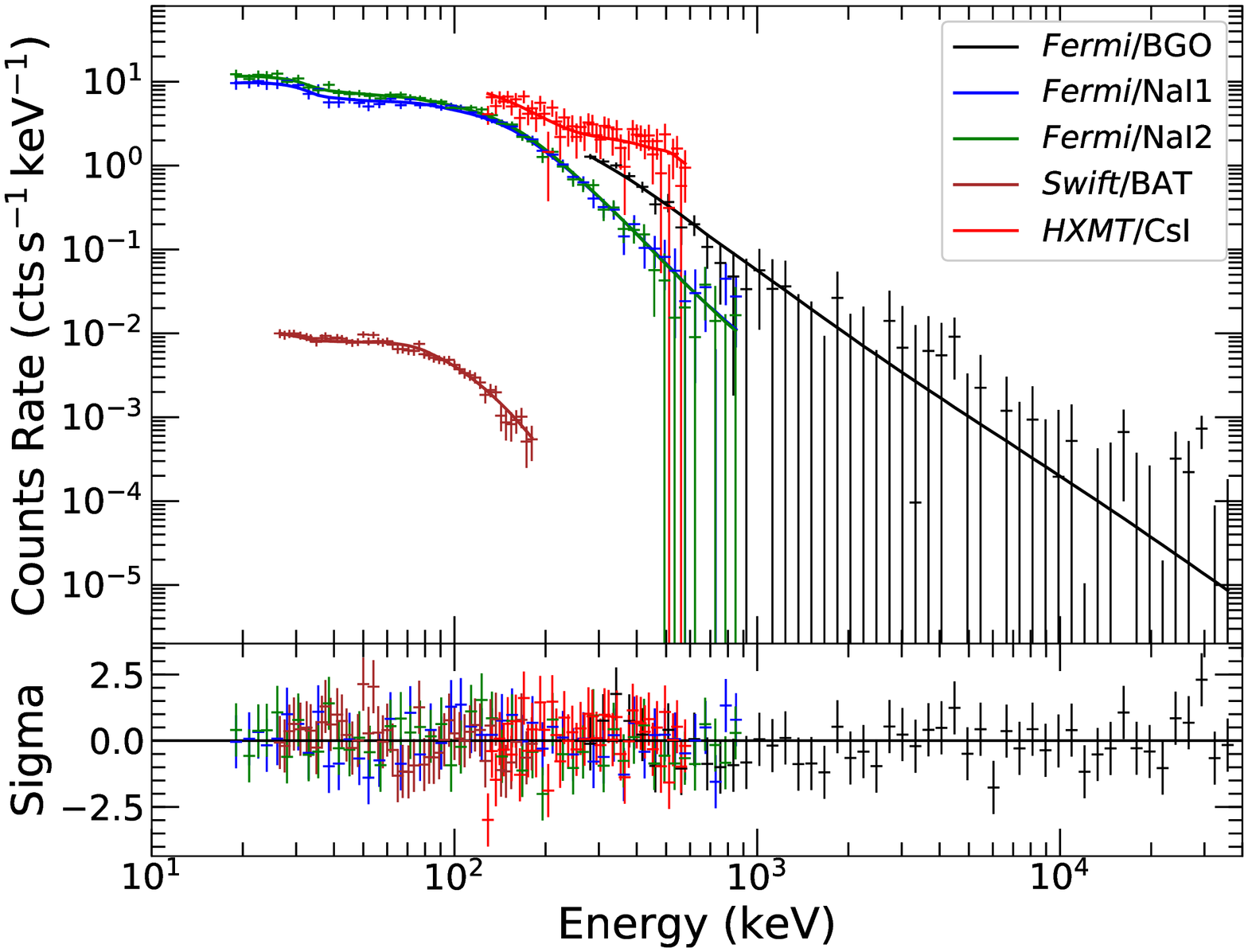}
    \end{minipage}%
    }%
    \quad
    \subfigure[GRB~181028A, LG mode]{
    \begin{minipage}[t]{0.5\linewidth}
    \centering
    \includegraphics[width=2.5in]{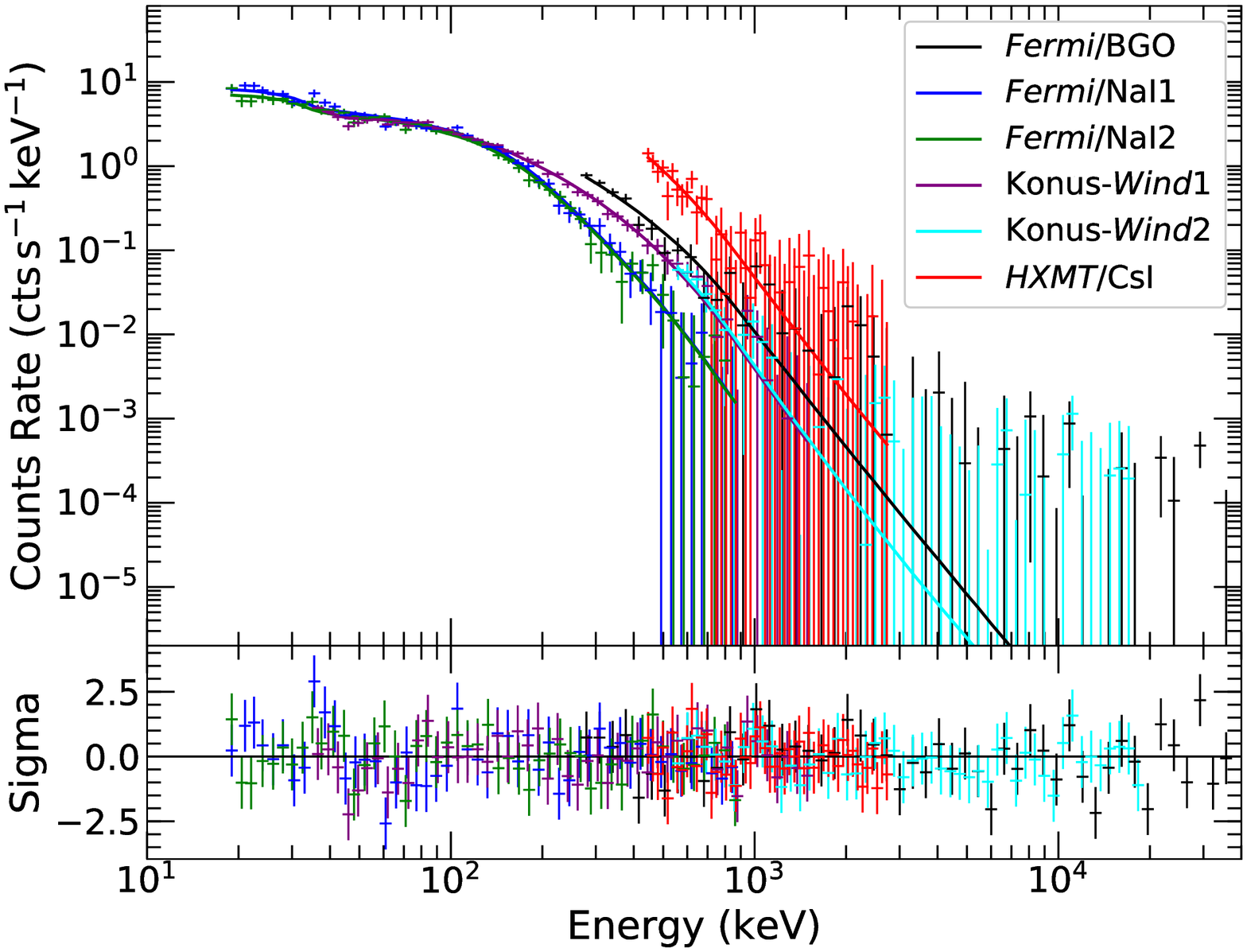}
    \end{minipage}%
    }%
    \subfigure[GRB~181212A, NG mode]{
    \begin{minipage}[t]{0.5\linewidth}
    \centering
    \includegraphics[width=2.5in]{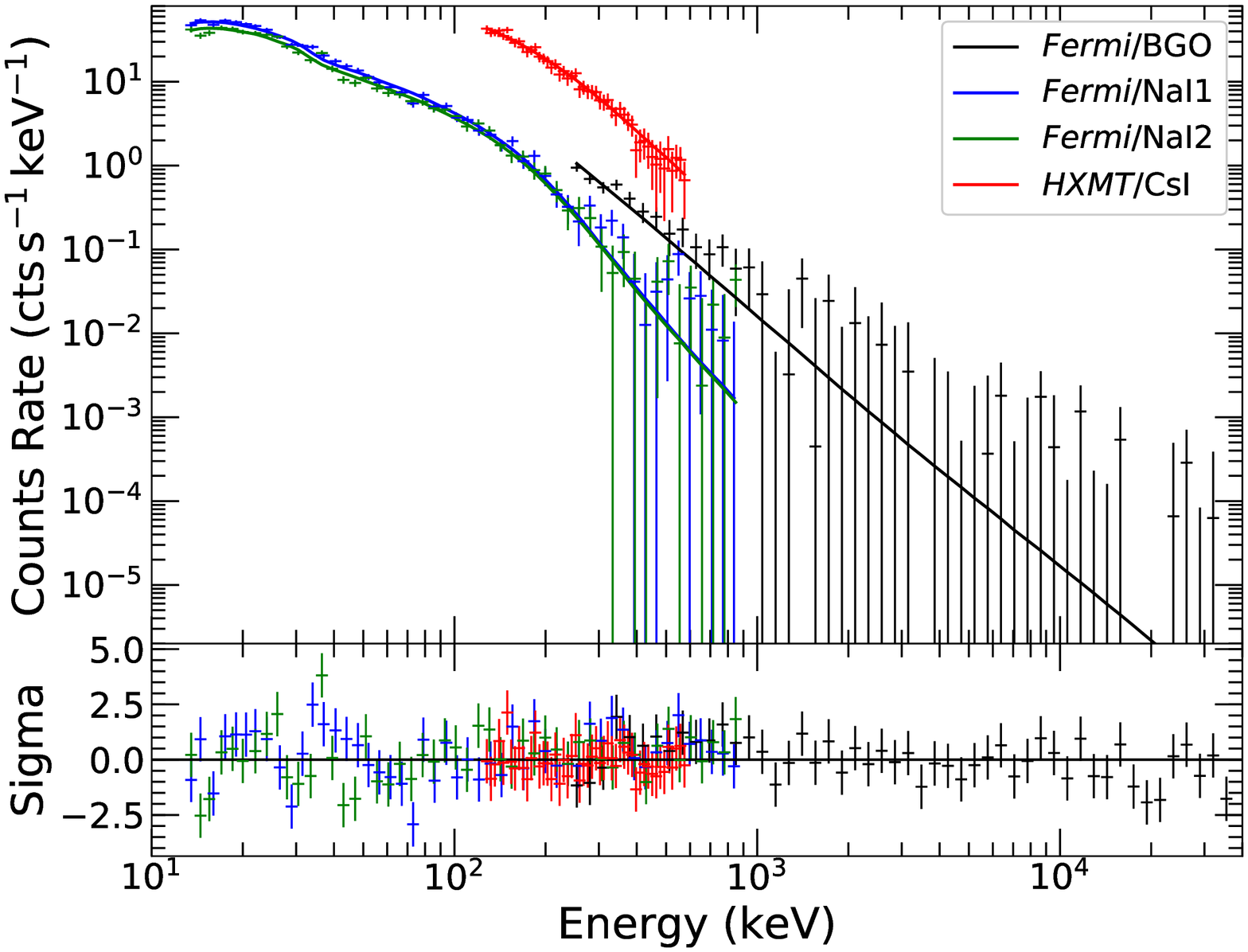}
    \end{minipage}%
    }%
    \caption{Joint spectral fitting of \emph{HXMT}/CsI (red), \emph{Fermi}/GBM BGO detectors (black), \emph{Fermi}/GBM NaI detectors (blue and green),
	\emph{Swift}/BAT (brown) and Konus-\emph{Wind} (purple and cyan).
	In the joint fittings, the 18 \emph{HXMT}/CsI spectra are merged and the merged spectrum are re-grouped to 50 energy bins for display clearly.}\label{10jpg}
    \end{figure*}
    \addtocounter{figure}{-1}
    \begin{figure*}[t]
    \addtocounter{subfigure}{6}
    \subfigure[GRB~190131A, NG mode]{
    \begin{minipage}[t]{0.5\linewidth}
    \centering
    \includegraphics[width=2.5in]{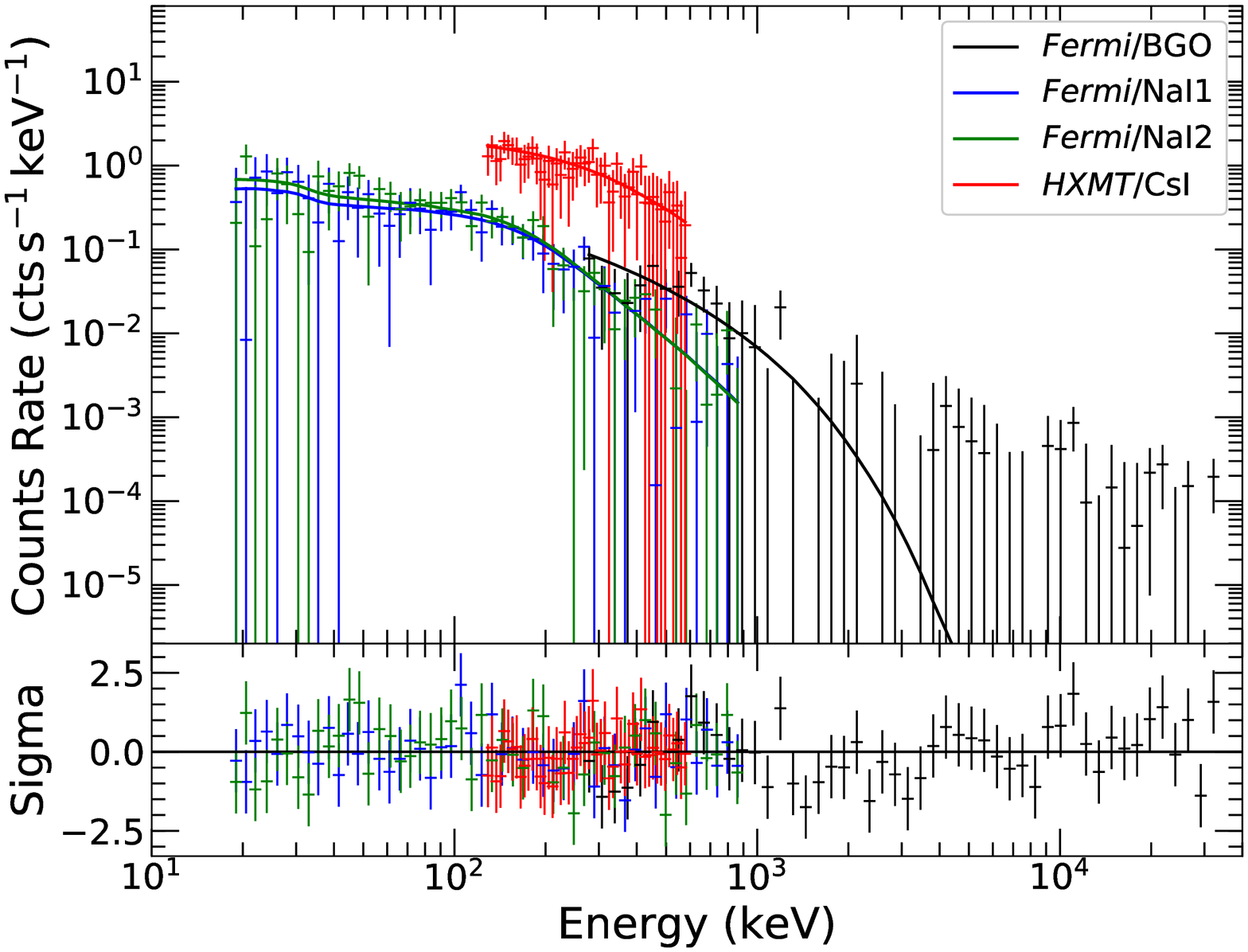}
    \end{minipage}%
    }%
    \subfigure[GRB~190324A, NG mode]{
    \begin{minipage}[t]{0.5\linewidth}
    \centering
    \includegraphics[width=2.5in]{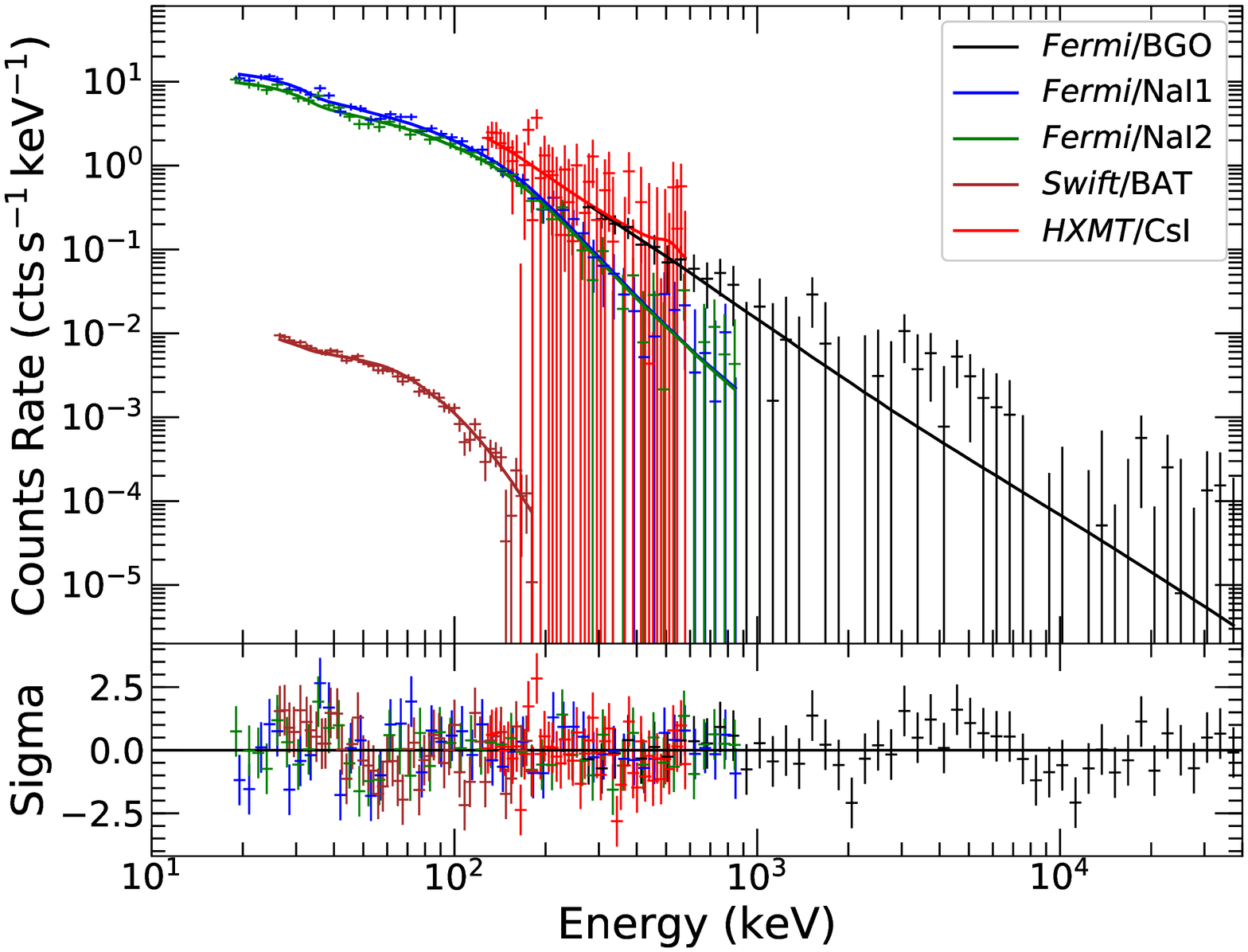}
    \end{minipage}%
    }%
    \quad
    \subfigure[GRB~190324B, NG mode]{
    \begin{minipage}[t]{0.5\linewidth}
    \centering
    \includegraphics[width=2.5in]{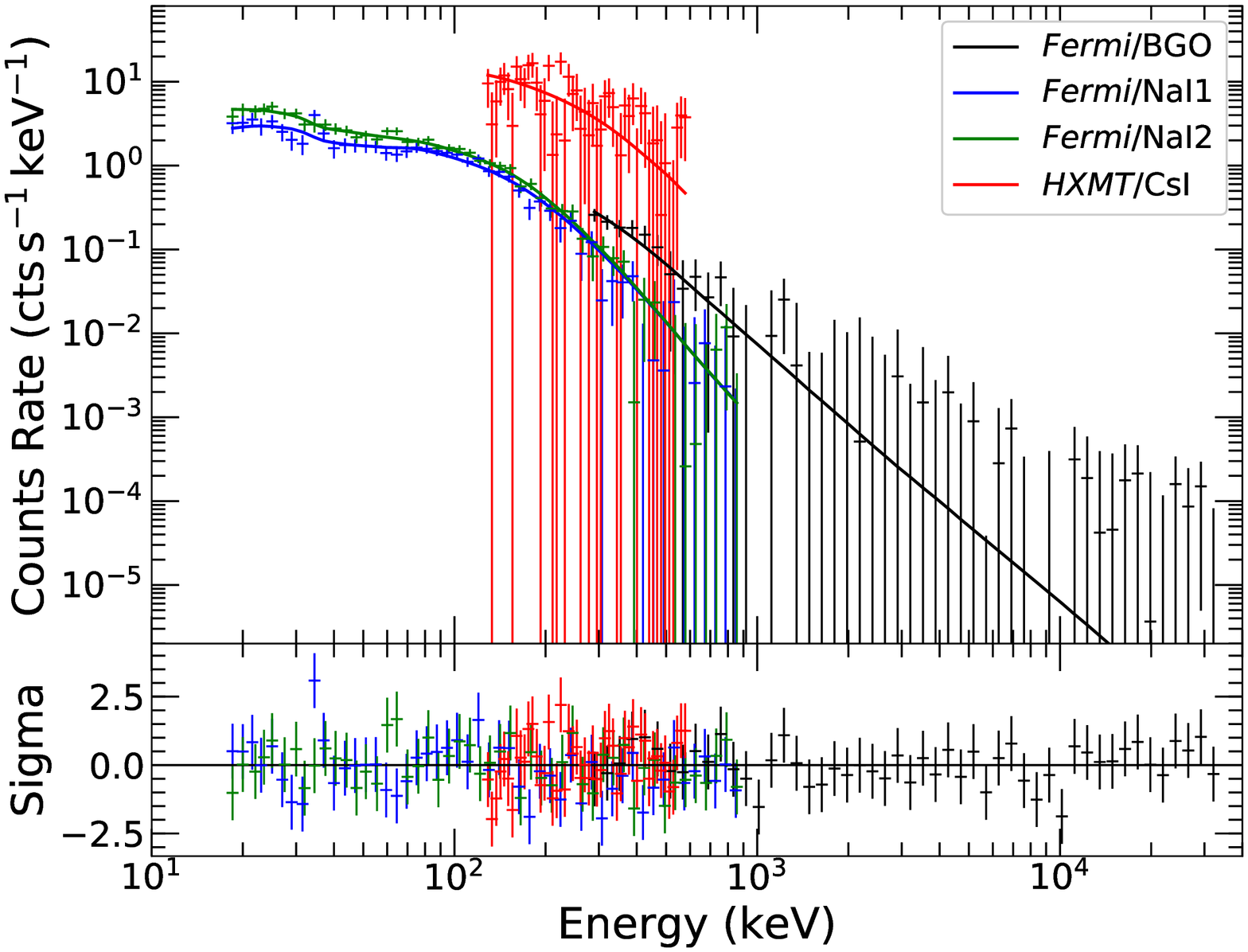}
    \end{minipage}%
    }%
    \subfigure[GRB~190326A, NG mode]{
    \begin{minipage}[t]{0.5\linewidth}
    \centering
    \includegraphics[width=2.5in]{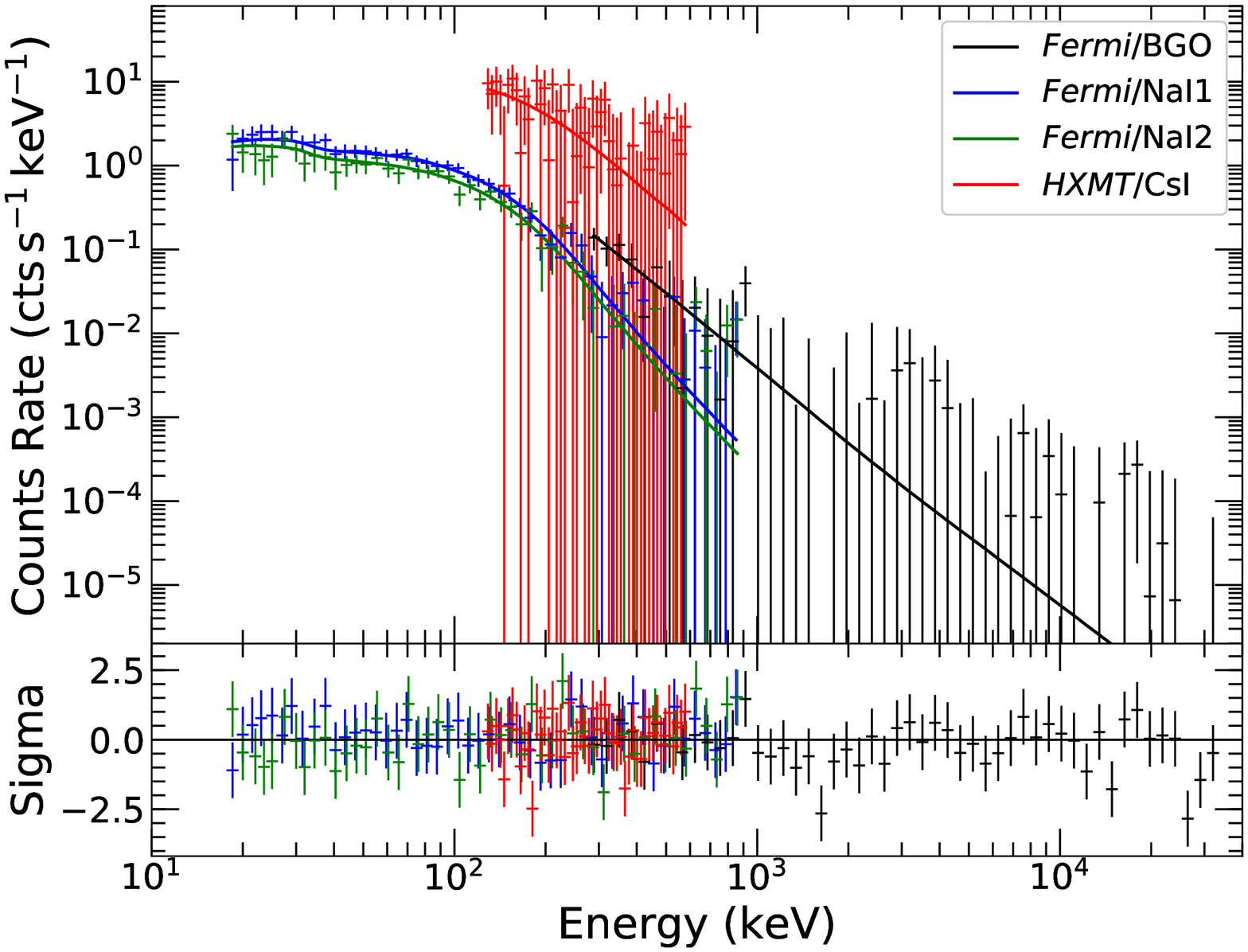}
    \end{minipage}%
    }%
    \centering
    \caption{(Continued)}
\end{figure*}

\section{Discussion and Conclusion}
\label{sec:Dis_Con}
The complicated \emph{Insight-HXMT} payload and platform can be simplified into a mass model,
which acts as the essential input for understanding the in-orbit behavior of the telescope via simulations.
Such a mass model can be tested by observing the Crab pulsed emission.
By comparing the Crab pulsed count rate ratio between the observed and the simulated,
we find that the ratio distribution has intrinsic dispersion $S_{\rm u}$ smaller than 0.3 and the ratio is close to
unit for photons at incident angle $\theta<90^\circ$. However, at other incident angles, the ratio sometimes tends to be larger,
probably due to the insufficiency in the built mass model.

With the blank sky and earth occultation observations,
the E-C relationship and the energy resolution of both the NG and LG modes are calibrated. We find that each
\emph{HXMT}/CsI detector has distinguish E-C relationship and evolves with time. Therefore, the E-C relationship is calibrated once per month.
The energy resolutions of all \emph{HXMT}/CsI detectors are stable and manifest with energy in an exponential manner.

The simulative spectral analyses with various spectral parameters and exposures are performed to demonstrate the capability of \emph{HXMT}/CsI in measuring the GRB spectrum.
Since the effective area is relatively small at below 100~keV, it is difficult to constrain the entire GRB spectral parameters with \emph{HXMT}/CsI alone.
However, with the large effective area around 1~MeV, the high-energy spectral index $\beta$ can be better constrained by Insight-HXMT/CsI with joint observations from other missions.
A systematic investigation upon the GRB flux in a maximum likelihood approach
show that the GRB flux given by \emph{HXMT}/CsI is systematically
higher by $7.0\pm8.8\%$ than that given by the other instruments. The difference is not significant and with a large statistical error due to the small number
of the sample, which also indicate the \emph{HXMT}/CsI instrumental response has been well calibrated.

For the important gamma-ray events with an accurate position, the incident direction in the payload coordinate system is known from
the satellite attitude. The instrumental response of this incident direction can also be calibrated specially with the flux of the
Crab pulse component, which can be considered as a standard manner for calibrating
\emph{Insight-HXMT} at gamma-rays.

In summary, the instrumental responses of the \emph{HXMT}/CsI detectors of \emph{Insight-HXMT} are well calibrated in aspects of mass model,
E-C relationship and the energy resolutions for both the NG and LG modes.
Thanks to the large effective area in the high-energy band, \emph{HXMT}/CsI shows its power in constraining the GRB spectrum
together with the campaigns with other missions which provide observations at lower energies.
We note that the current detection efficiency with the
incident angle $\theta>90^\circ$ may be somewhat overestimated.
The response with the incident angle $\theta>90^\circ$ will be improved empirically in future with bright GRB campaigns and more Crab observations.

\section*{Acknowledgements}
This work made use of the data from the \emph{Insight-HXMT} mission, a project funded by China National Space Administration (CNSA)
and the Chinese Academy of Sciences (CAS). The authors thank supports from the National Program on Key Research and Development Project
(Grants No. 2016YFA0400802, 2016YFF0200802), the National Natural Science Foundation of China under Grants No. U1838202, U1838201, U1838110, U1838113 and U1838104,
and the Strategic Priority Research Program on Space Science, the Chinese Academy of Sciences, Grant No. XDB23040400.
The authors are grateful to Dmitry Frederiks and the KW team for providing the KW data and their efforts to perform the joint fitting.
We thank Kazutaka Yamaoka for his helpful discussions.

\section*{References}
\bibliographystyle{aasjournal}
\bibliography{mybib}
\end{document}